\documentclass{kapmono} 
\usepackage{epsfig}

\usepackage{monops} 

\normallatexbib

\begin{document}

\chapter[]
{Luttinger liquids: The basic concepts}






\noindent {\large K.Sch\"onhammer}

\noindent Institut f\"ur Theoretische Physik, Universit\"at
  G\"ottingen, Bunsenstr. 9, D-37073 G\"ottingen, Germany\\\

\noindent{\bf Abstract}

 \noindent This chapter reviews the theoretical description
of interacting fermions in one dimension.
The Luttinger liquid concept is elucidated
using the Tomonaga-Luttinger model as well as integrable lattice models.
Weakly coupled chains and the attempts to experimentally verify the   
theoretical predictions are discussed.\\\

\noindent{\bf Keywords}

\noindent Luttinger liquids, Tomonaga model, bosonization,
anomalous power laws, breakdown of Fermi liquid theory, spin-charge
separation, spectral functions, coupled chains,
quasi-one-dimensional conductors

\vspace{1.5cm}

\section{  Introduction} 

In this chapter we attempt  a simple selfcontained introduction
to the main ideas and important computational tools 
for the description of interacting fermi\-ons in one spatial dimension.
The reader is expected to have some knowledge of the method
of second quantization. As in section 3 we describe a {\it constructive}
approach to the important concept of bosonization, {\it no} 
quantum field-theoretical background is required.
After mainly focusing on the Tomonaga-Luttinger model in 
sections 2 and 3 we present results for integrable lattice models in section 4.
In order to make contact to more realistic systems the coupling
of strictly $1d$ systems as well as to the surrounding is 
addressed in section 5. 
 The attempts to
experimentally verify typical Luttinger liquid features like
anomalous power laws in various correlation functions are only
shortly discussed as this is treated in other chapters of this book. 

\section{  Luttinger liquids - a short history of the ideas} 
As an introduction the basic steps towards the general concept of
Luttinger liquids are presented in historical order. In this
exposition the ideas are discussed without presenting all
technical details. This is done in section 3 by disregarding the
historical aspects aiming at a simple presentation of the important
practical concepts like the `` bosonization of field operators''.
\subsection{ Bloch's method of ``sound waves'' (1934)}
In a paper on incoherent x-ray diffraction Bloch \cite{Bloch} realized and
used the fact that one-dimensional $(d=1)$ noninteracting fermions
have the same type of low energy excitations as a harmonic chain.
The following discussion of this connection is very different
from Bloch's  own presentation.
 
\noindent The low energy excitations
determine  e.g. the low temperature  
specific heat. Debye's famous $T^3$-law
for the lattice contribution of three dimensional solids reads in $d=1$
\begin{equation}
c_L^{\rm Debye} = \frac{\pi}{3}k_B \left(\frac{k_BT}{\hbar c_s}\right),
\label{Debye}
\end{equation}
where $c_s$ is the sound velocity. At low temperatures
 the electronic contribution to the
specific heat in the ``Fermi gas'' approximation of Pauli is also
linear in $T$ and involves the density of states
of the
noninteraction electrons at the Fermi energy. This yields for spinless
fermions in $d=1$

\begin{equation}
c_L^{\rm Pauli}= \frac{\pi}{3}k_B \left(\frac{k_BT}{\hbar v_F}\right),
\label{Pauli}
\end{equation}
where $v_F$ is the Fermi velocity. With the replacement $c_s
\leftrightarrow v_F$ the results are {\it identical}. This suggests
that apart from a scale factor the (low energy) excitation energies
and the degeneracies in the two types of systems
 are identical. For the harmonic chain the
excited states are classified by the numbers ${n_k}$ of phonons in the
modes $\omega_k$ whith $n_k$ taking integer values from zero to
infinity. The excitation energy is given by $E(\{n_{k_{m}}\}) - E_0 =
\sum_{k_{m}}\hbar \omega_{k_{m}}n_{k_{m}}$. For small wave numbers
$k_m$ the dispersion is {\it linear} $\omega_{k_{m}}
\approx c_s|k_m|$. Therefore the excitations energies are 
  multiples of $\hbar c_s(2\pi/L)$ for periodic boundary
conditions and multiples of 
$\Delta_B \equiv \hbar c_s \pi/L$ for fixed boundary
conditions. The calculation of the partition function is standard
textbook material. This is also true for noninteracting electrons but
there the calculation involves fermionic occupation numbers $n^F_k$
which take values zero and one. The two textbook calculations
yield Eqs.\ (\ref{Debye}) and (\ref{Pauli}), but through the ``clever''
use
 of the grand
canonical ensemble in order to simplify
 the fermionic calculation the identity (apart
from $c_s \leftrightarrow v_F$) remains mysterious. A deeper
understanding involves two steps:

1) {\it Linearization} of the kinetic energy $\varepsilon_k =
  \hbar^2 k^2/(2m)$ of the free fermions around the Fermi point $k_F$
  for fixed boundary conditions
or both Fermi points $\pm k_F$ for periodic boundary conditions.
 As the argument is simplest for fixed
  boundary conditions \cite {SM} which lead to $k_m = m \pi/L$ we discuss
  this case for the moment. Then the energies $\varepsilon_{k_{n}} -
  \varepsilon_F$ are integer multiples of $\Delta_F \equiv \hbar v_F
  \pi/L$ where $v_F$ is the Fermi velocity.

\begin{figure}[hbt]
\begin{center}
\epsfxsize4.0cm
\epsffile{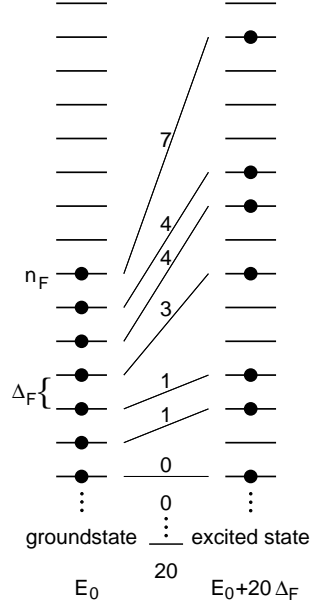}
\end{center}
\caption{Example for the classification scheme for the excited states
  in terms of the numbers $n_j$ of upward shifts by $j$ units of
  $\Delta_F$. In the example shown
the nonzero $n_j$ are $n_7=1, n_4=2, n_3=1$ and $ n_1=2$.}
\label{Fig.1}
\end{figure}

 2) {\it Classification} of any state of the Fermi system by
  the {\it number} $n_j$ {\it
of upward shifts by} $j$  {\it units of} $\Delta_F$
with respect to the ground state. As the
  fermions are indistinguishable the construction of the $\{n_j\}$
  shown in Fig. \ref{Fig.1}, where the highest occupied level in the excited
  state is connected with the highest occupied level in the ground
  state and so forth for the second, third \ldots highest levels,
  completely specifies the excited state. As the $n_j$ can run from
  zero to infinity like bosonic quantum numbers and the excitation
  energy is given by $\sum_j(j \Delta_F)n_j$ the {\it canonical}
  partition function for the noninteracting fermions has the same form
  as the canonical partition function for the harmonic chain apart from
  $\Delta_F \leftrightarrow \Delta_B$ if one assumes the Fermi sea to
  be infinitely deep \cite {Fermisee}.

\noindent As we have linearized $\omega_k$
  for small $k$ as well as $\varepsilon_k$ around $k_F$ this
  equivalence only holds for the {\it low temperature} specific heats
  $(k_BT\ll \hbar \omega_{\rm max},k_BT \ll \varepsilon_F)$.

 If we denote the creation (annihilation) operator of a fermion with
$k_n = n\pi/L$ by $c^\dagger_n (c_n)$ and assume a {\it strictly linear}
dispersion $\varepsilon_n^{\rm lin} = \hbar v_F k_n$
for {\it all} $k_n>0$ a more technical
formulation of the discussed equivalence can be given by the 
{\it exact operator identity}
\begin{eqnarray}
T & = & 
\sum^\infty_{n=1}\hbar v_F k_n c^\dagger_n c_n \nonumber\\
&=&
\frac{\hbar v_F\pi}{L}
 \left[ \sum^\infty_{l=1} l\; b^\dagger_l\; b_l + \frac{1}{2}
{\cal N} ({\cal N} +1)\right],
\label{Kronig}
\end{eqnarray}
where the operators $b_l$ with $l \ge 1$ are defined as
\begin{equation}
b_l \equiv \frac{1}{\sqrt{l}} \sum^\infty_{m=1}c^\dagger_m c_{m+l}
\label{bl}
\end{equation}
and ${\cal N} \equiv \sum^\infty_{n=1}c^\dagger_n c_n$ is the
fermionic particle number operator. The proof of 
the ``Kronig identity'' (\ref{Kronig}) is simple
(see Ref. {\cite S}) . The operators $b_l$ obey commutation
relations $[b_l,b_{l'}] = 0$ and for $l \ge l'$
\begin{equation}
\left[b_l,b^\dagger_{l'}\right] = \frac{1}{\sqrt{ll'}} \sum^{l'}_{m=1}
c^\dagger_{m}c_{m+l-l'}.
\label{Komm1} 
\end{equation}
For all $N$-particle states $|\phi^{(M)}_N\rangle \equiv \prod^N_{n=1}
c^\dagger_{i_{n}}|\mbox{Vac}\rangle$ in which the $M(<N)$ lowest one-particle
levels are {\it all occupied} one obtains
\begin{equation}
\left[b_l,b^\dagger_{l'}\right] |\phi^{(M)}_N\rangle =
\delta_{ll'}|\phi^{(M)}_N\rangle
\label{Komm2} 
\end{equation}
for $l,l' \le M$, i.e. these operators obey {\it boson commutation
  relations}  $[b_l,b^\dagger_{l'}] =\delta_{ll'}\hat 1 $
 in this {\it subspace} of all possible $N$-particle
states.

\noindent Later it turns out to be useful to work with
$\tilde T\equiv T-\langle T\rangle_0-\mu_0 \tilde {\cal N}$, where
$\langle T\rangle_0=\Delta_Fn_F(n_F+1)/2 $ is the ground-state energy,
$\mu_0=\Delta_F(n_F+1/2)$ is the chemical potential of the
noninteracting fermions and  $\tilde {\cal N}\equiv 
{\cal N}-n_F\hat 1 $.
  Then $\tilde T $ is of the form as the rhs of Eq.\ (\ref{Kronig}) with
 ${\cal N} ({\cal N} +1) $ 
 replaced by $\tilde {\cal N}^2$.

\subsection{Tomonaga (1950): Bloch's method of sound waves applied to
   interacting fermions}
When a two-body interaction between the fermions is switched on, the
ground state is no longer the filled Fermi
sea but it has admixtures of (multiple) particle-hole pair
excitations. In order to simplify the problem Tomonaga studied the
{\it high density limit} where the range of the interaction is much
larger than the interparticle distance, 
using {\it periodic} boundary conditions \cite {T}. Then the Fourier transform
$\tilde v(k)$ of the two-body interaction is nonzero only for values
$|k| \le k_c$ where the cut-off $k_c$ is much smaller than the Fermi
momentum $ k_c \ll k_F $.
This implies that for not too strong interaction the ground state and
low energy excited states have negligible admixtures of holes deep in
the Fermi sea and particles with momenta $|k|-k_F\gg k_c$ .
 In the two intermediate regions around the two Fermi points $\pm
 k_F$, with
particle-hole pairs present, the dispersion
$\varepsilon_k$ is {\it linearized} in order to apply Bloch's
``sound wave method'' 
\begin{equation}
k \approx \pm k_F : \quad \varepsilon_k = \varepsilon_F \pm v_F (k\mp
k_F).
\label{lin}
\end{equation}
Tomonaga realized that the Fourier components of the operator of the
density 
\begin{eqnarray}
\hat \rho_n &=&
 \int^{L/2}_{-L/2} \hat \rho(x)e^{-ik_nx} dx =\int^{L/2}_{-L/2}
 \psi^\dagger (x) \psi (x) e^{-ik_nx} dx\nonumber \\
&=&
\sum_{n'} c^\dagger_{n'}c_{n'+n},
\label{Dichte}
\end{eqnarray}
where $c^\dagger_n (c_n)$ creates (annihilates) a fermion in the state
with momentum $k_n = \frac{2\pi}{L}n$, plays a central role in the
interaction term, as well as the kinetic energy.
Apart from an additional
term linear in the particle number operator \cite {S},
which is usually neglected, the two-body interaction is given by
\begin{equation}
 \hat V =
\frac{1}{2L}\sum_{n\ne 0} \tilde v(k_n)\hat\rho_n \hat\rho_{-n} +
\frac{1}{2L} {\cal N}^2 \tilde v(0)
\label{WW}
\end{equation}
 Tomonaga's important step
was to split $\hat \rho_n$ for $|k_n|\ll k_F$ into two parts, one
containing operators of {\it ``right movers''} i.e. involving fermions
near the right Fermi point $k_F$ with velocity $v_F$ and {\it ``left
  movers''} involving fermions near $-k_F$ with velocity $-v_F$
\begin{equation}
\hat \rho_n = \sum_{n'\ge 0}c^\dagger_{n'}c_{n'+n} +\sum_{n'<
  0} c^\dagger_{n'} c_{n'+n} \equiv \hat \rho_{n,+} + \hat \rho_{n,-}
\label{Zerlegung}
\end{equation}
where the details of the splitting for small $|n'|$ are irrelevant. Apart
from the square root factor the $\hat\rho_{n,\alpha}$ are
similar to the $b_l$ 
defined in Eq.\ (\ref{bl}) . Their commutation relations in the {\it low energy
subspace} are 
\begin{equation}
[\hat\rho_{m,\alpha}, \hat\rho_{n,\beta}] = \alpha m \delta_{\alpha \beta}
\delta_{m,-n} \hat 1.
\label{Komm3}
\end{equation}
If one defines the operators
\begin{eqnarray}
b_n \equiv \frac{1}{\sqrt{|n|}}\left\{ \begin{array}{ll}
\hat\rho_{n,+} \hspace*{3ex} & \mbox{ for } n > 0\\
\hat\rho_{n,-} & \mbox{ for }n  < 0\end{array} \right.
\label{Boson1}
\end{eqnarray}
and the corresponding adjoint operators $b^\dagger_n$   this leads  using
$\rho^\dagger_{n,\alpha} = \rho_{-n,\alpha}$ to the bosonic
commutation relations
\begin{equation}
[b_n, b_m] = 0, \quad [b_n,b^\dagger_m] = \delta_{mn} \hat 1.
\label{Komm4}
\end{equation}
Now the kinetic energy of the right movers as well as that of the left
movers can be ``bosonized'' as in Eq.\ (\ref{Kronig}).
The interaction $\hat V$ is bilinear in
the $\hat\rho_n$ as well as the $\hat\rho_{n,\alpha}$. 
  Therefore apart from an additional term
containing particle number operators
 {\it the Hamiltonian for the interacting fermions is a
  quadratic form in the boson operators}
\begin{eqnarray}
\tilde H & = &
 \sum_{n>0}\hbar k_n \left\{
\left(v_F + \frac{\tilde v(k_n)}{2 \pi\hbar }\right)  
\left(b^\dagger_n b_n + b^\dagger_{-n} b_{-n}\right) \right. \nonumber
\\ && \left. +
\frac{ \tilde v(k_n)}{2 \pi \hbar}
\left(b^\dagger_n b^\dagger_{-n} + b_{-n} b_n\right)\right\}
+ \frac{\hbar \pi}{2L}\left[v_N\tilde{\cal N}^2
 +v_J{\cal J}^2\right] \nonumber \\ & \equiv & H_B +
H_{\tilde{\cal N,}{\cal J}},
\label{T1}
\end{eqnarray}
where $\tilde{\cal N}\equiv \tilde{\cal N}_+ +\tilde {\cal N}_-$
 is the total particle
number operator relative to the Fermi sea,
${\cal J}\equiv \tilde{\cal N}_+ - \tilde{\cal N}_-$ the ``current operator'',
 and the
velocities are given by $v_N = v_F + \tilde v(0)/\pi\hbar$ and $v_J =v_F$.
 Here $v_N$ determines the energy change for adding particles
without generating bosons while
$v_J$ enters the energy change when the difference in the number of
right and left movers is changed. Similar to the discussion at the end
of section 1.1 we have defined $\tilde H \equiv 
H-E_0^H -(\mu_0+\tilde
v(0)n_0)\tilde {\cal N}$, where
$ E_0^H  $ is the Hartree energy and
 $n_0$  the particle density.
 As the particle number operators
$\tilde{\cal N}_\alpha$ commute with the boson operators $b_m(b^\dagger_m)$
the two terms $H_B$ and $H_{\tilde{\cal N},{\cal J}}$ in the
 Hamiltonian {\it commute} and
can be treated separately.
Because of the translational invariance the two-body interaction
only couples the modes described by $b^\dagger_n$ and $b_{-n}$.
 With the Bogoliubov transformation
  $\alpha^\dagger_n = c_n b^\dagger_n - s_n
b_{-n}$
the Hamiltonian $H_B$
can be brought into the form

\begin{equation}
H_B = \sum_{n \neq 0}\hbar  \omega_n \alpha^\dagger_n \alpha_n + \;
\mbox{const}.,
\label{T2}
\end{equation}
where the  $\omega_n = v_F |k_n|\sqrt{1+\tilde v(k_n)/\pi\hbar v_F}$
follow from $2\times 2$ eigenvalue
problems corresponding to the condition
$[H_B,\alpha^\dagger_n]=\hbar\omega_n
\alpha^\dagger_n$.
         For small $k_n$ one obtains
for smooth potentials $\tilde v(k)$ again a {\it
  linear} dispersion $\omega_n\approx v_c|k_n|$, with the
{\it ``charge velocity''} $v_c = \sqrt{v_N v_J}$, which is larger than
$v_F$ for
  $\tilde v(0) >0$     . The expression for the coefficients
$c_n$ and $s_n$ with $c^2_n - s^2_n = 1$  will be presented later
for the generalized model Eq.\ (\ref{TL1}) . For fixed particle numbers
$N_+$ and $N_-$, the excitation energies of the {\it interacting}
Fermi system are given by $\sum_m \hbar \omega_m n_m$ with integer
occupation numbers $0\le n_m<\infty$.
For small enough excitation energies the only difference 
of the excitation spectrum 
for fixed particle numbers with respect to the noninteracting case is 
the replacement $v_F\leftrightarrow v_c$.

 In his seminal paper Tomonaga did not realize
the anomalous decay of correlation functions of the model
because in his discussion of the density correlation function
he missed the $2k_F$-contribution discussed in section 3.
\begin{figure} [hbt]
\begin{center}
\epsfig{file=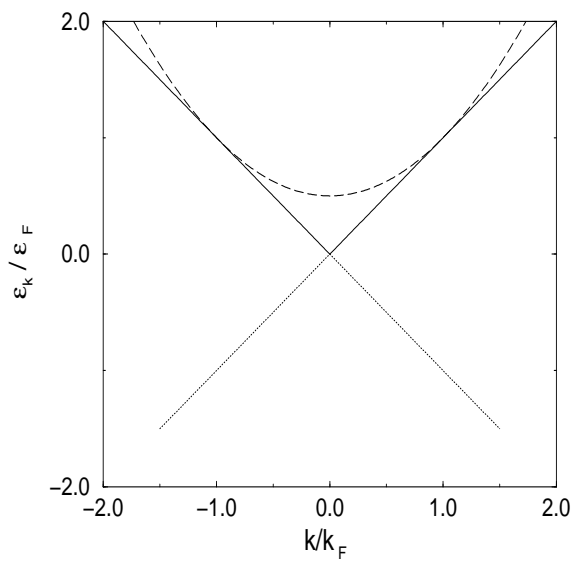,width=7.5cm,height=7.5cm,angle=-90}
\caption{Energy dispersion as a function of momentum. The dashed curve
shows the usual ``nonrelativistic'' dispersion and the full curve the
 linearized version used (apart from a constant shift) in
Eq.\ (\ref{Kronig}) for $k>0$ for fixed boundary conditions.
The dot-dashed parts are the additional states for $k_0=-1.5k_F$. The 
model discussed by Luttinger corresponds to $k_0 \to -\infty$.}
\end{center}
\label{Fig.2}
\end{figure}

\subsection{Luttinger (1963): no discontinuity at the Fermi surface}
Luttinger, apparently unaware of Tomonaga's work,
 treated spinless, massless fermions (in the relativistic sense,
but $c \leftrightarrow v_F$) in one dimension, i.e. two
{\it infinite} branches
of right and left moving fermions with dispersion $\pm v_F k$
\cite {L}. As
Luttinger himself made an error with the fact that his
Hamiltonian is not bounded from below, it is useful to switch from
Tomonaga's to Luttinger's model keeping a {\it band cut-off} $k_0$ such that 
$k\ge k_0=2\pi m_0/L $ with $m_0<0$ for the right movers 
and correspondingly for the left movers (see Fig. \ref{Fig.2}).
Fortunately Luttinger's error
had no influence on his inquiry if a {\it sharp Fermi surface} exists
in the exact ground state of the interacting model. After a rather
complicated calculation using properties of so-called ``Toeplitz
determinants'' Luttinger found that the average occupation
 $\langle n_{k,+} \rangle$ in
the ground state for $k \approx k_F$ behaves as
\begin{equation}
\langle n_{k,+} \rangle -\frac{1}{2} \sim \left|\frac{k - k_F}{k_c}
\right|^{\alpha_L}
\mbox{sign} {(k_F-k)},
\label{nvk}
\end{equation}
where $\alpha_L$ depends on the interaction strength (see
below) \cite {Kommentar}.
 ``Thus, in this model, the smallest amount of interaction
{\it always} destroys the discontinuity of $\langle n_k\rangle$
at the Fermi surface'' \cite {L}.
This can be related to the fact that
the equal time correlation functions
$\langle\psi^\dagger_\alpha(x)\psi_\alpha(0)\rangle$ decay
 as $1/|x|^{1+\alpha_L}$
in the {\it interacting} system in contrast to
$\langle\psi^\dagger_\alpha(x)\psi_\alpha(0)\rangle \sim 1/|x|^d$
 (with $d = 1$) in the
noninteracting case. Therefore $\alpha_L$ is called the {\it ``anomalous
  dimension''}\cite {Index}.

Apart from the different dispersion Luttinger also
used a different interaction. In contrast to
Tomonaga he only kept an interaction between the right and left movers
but not the term $\sim\tilde v(k_n)(b^\dagger_n b_n + b^\dagger_{-n}
b_{-n})$ in Eq.\ (\ref{T1}) . In the limit of a delta interaction of the right
and left movers his model is identical to the {\it massless Thirring
  model} (1958) \cite {Thirring} at that 
time not well known in the solid state physics
community.

\subsection{Towards the ``Luttinger liquid'' concept}
\begin{figure} [hbt]
\begin{center}
\epsfig{file=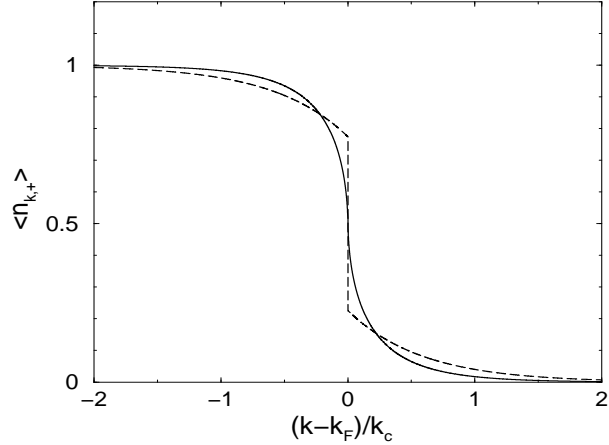,width=6cm,height=8cm,angle=-90}
\caption{ The full line 
shows the average occupation $\langle n_{k,+} \rangle $ for a
TL model with 
$\alpha_L=0.6$. The dashed line shows the expectation from Fermi
liquid theory, where the discontinuity at $k_F$ determines
the quasi-particle weight $Z_{k_F}$ in $\rho_+(k_F,\omega)  $.
As discussed following Eq.\ (\ref{anomaleD}) this can also be realized in a TL
model with $g_2(0)=0$.
 There also the details of the
interaction are specified.}
\end{center}
\label{Fig.Lutt}
\end{figure}

Luttinger's treatment of the Dirac sea 
 was corrected in a paper by Mattis and Lieb (1965)
\cite {ML} which also offered a simpler way to calculate $\langle
n_{k,\alpha}\rangle$. The time dependent one-particle Green's function for the
spinless Luttinger model was calculated by Theumann (1967) \cite {Th}
by generalizing this method. She found {\it power law behaviour} in
the corresponding 
spectral function $\rho(k,\omega)$, especially
$\rho(k_F,\omega)\sim \alpha_L|\omega|^{\alpha_L-1}$, i.e. {\it no sharp
  quasiparticle for $k = k_F$}
consistent with Luttinger's result 
for the occupation numbers (Fig.\ref{Fig.Lutt}).
 For a delta interaction her results
    agreed with an earlier calculation for the massless Thirring model
    by Johnson (1961) \cite {Johnson}.
 Later the time dependent one-particle Green's
    function was calculated by various other methods, e.g. using Ward
    identities (Dzylaloshinski and Larkin (1974) \cite {DL}) as well as the
    important method of the {\it''bosonization of the field
      operator''} (Luther and Peschel (1974) \cite {LP})
 which will be addressed
    in detail in section 3. It was first proposed in a different
    context by Schotte and Schotte (1969) \cite {SS}.

\noindent What is now usually called the ``Tomonaga-Luttinger (TL) model'' is
the following generalization of Eq.\ (\ref{T1}) 
\begin{eqnarray}
\hspace{-0.5cm} \tilde H_{TL} & = & \frac{2\pi\hbar}{L}
\sum_{n > 0}n \left\{
\left ( v_F + \frac{g_4(k_n)}{2 \pi\hbar}\right )\left( b^\dagger_n b_n +
    b^\dagger_{-n}b_{-n}\right)\right. \nonumber
\\  
\hspace{-0.5cm} & +& \left.
\frac{g_2(k_n)}{2\pi\hbar}\left(b^\dagger_n b^\dagger_{-n}+b_{-n} b_n\right)
\right\}
 + \frac{\hbar \pi}{2L}\left\{v_N \tilde {\cal N}^2 +
 v_J {\cal J}^2\right\},
\label{TL1}
\end{eqnarray}
where $v_N=v_F +(g_4(0)+g_2(0))/2\pi\hbar$ and 
$v_J=v_F +(g_4(0)-g_2(0))/2\pi\hbar$.
The interaction parameters $g_2(k_n)$ and  $g_4(k_n)$
are allowed to be {\it different}.
 As Tomonaga's  original  model the TL model is exactly solvable,
i.e. it can also be brought into the form of Eq.\ (\ref{T2}).
The eigenvector components in $\alpha^\dagger_n=c_n b^\dagger_n-s_n
b_{-n}$
are given by
\begin{equation}
c_n=\frac{1}{2}\left (\sqrt{K_n}+\frac{1}{\sqrt{K_n}}\right), \quad
s_n=\frac{1}{2}\left (\sqrt{K_n}-\frac{1}{\sqrt{K_n}}\right)
\label{Koeff}
\end{equation}
with $K_n=\sqrt{ v_J(k_n)/v_N(k_n)}$, where $v_{J(N)}(k_n)\equiv 
v_F  +[g_4(k_n)\mp g_2(k_n)]/2\pi\hbar $.
The frequencies are given by $\omega_n=|k_n| \sqrt{ v_J(k_n)v_N(k_n)}
\equiv |k_n|v_c(k_n) $.
\noindent  The TL-Hamiltonian corresponds to a 
 fermionic Hamiltonian  that {\it conserves} the number of right and
 left movers.

A more general model
including spin and 
terms changing right movers into left movers and vice versa
 is usually called the ``$g$-ology
model''. An important step towards the general Luttinger liquid
concept came from the {\it renormalization group} (RG) study of this
model. It was shown that for {\it repulsive}
interactions  (see section 3) the renormalized interactions flow
towards a fixed point Hamiltonian of the TL-type
unless in lattice models for commensurate electron fillings
strong enough interactions (for the half filled 
Hubbard model discussed in section $4$ this happens for
arbitrarily small on-site  Coulomb interaction U) destroy the
metallic state by opening a Mott-Hubbard gap. The RG approach
is described in
detail in reviews by S{\'o}lyom (1979) \cite {So}
and Shankar (1994) \cite {Shankar}.
 These results as well as
insight from models which allow an exact solution by the Bethe
  ansatz led Haldane \cite {H1,H}
 to propose the concept of {\it Luttinger
  liquids} (LL) as a replacement of Fermi liquid theory in one
dimension, which ``fails because 
 of the infrared divergence of certain vertices it assumes to remain 
finite'' \cite{H} .
 At least for
spinless fermions Haldane was able to show that ``the Bogoliubov
transformation technique that solves the Luttinger model provides a
general method for resumming {\it all} infrared divergences
present''\cite {H}.
 Similar to Fermi liquid theory in higher dimensions this new
LL phenomenology allows to describe the low energy physics in terms of
a few constants, {\it two} for the spinless case: the ``stiffness constant''
  $K \equiv K_0=\sqrt{v_J/v_N}$ (also called $g$
in various publications) and the {\it ``charge velocity''}
$v_c=\sqrt{v_J v_N}$. In his seminal paper Haldane
showed explicitly that the LL relations survive in not
exactly soluble 
generalizations of the TL model with a non-linear fermion dispersion.
He also gave a
clear presentation how to calculate general correlation functions
and e.g. the occupancies
shown in Fig. \ref{Fig.Lutt} for the TL model. The
technical details are addressed in section 3.

Before we do this two additional important aspects of LL-behaviour should be 
mentioned. The first concerns the strong influence of impurities
on the low energy physics
\cite {GoDa,Mattis,LuPe,ApRi,GiaS,KF}, especially the
 peculiar modification of the
electronic properties of a LL when a {\it single} impurity with
an arbitrarily weak backscattering potential is present. For a 
spinless LL with a {\it repulsive } two-body interaction, i.e. $K<1$
a perturbative bosonic RG calculation \cite {KF} shows that 
 the backscattering strength $V_B$ is a {\it relevant}
perturbation which grows as $\Lambda^{K-1}$ when the flow parameter $\Lambda$
is sent to zero. This leads to a breakdown of the perturbative
analysis in $V_B$. On the other hand a weak hopping between the open
ends of two semi-infinite chains is {\it irrelevant} and scales to
zero as $\Lambda^{K^{-1}-1}$. Assuming that the open chain presents
the only stable fixed point it was argued that at low energy scales 
even for a weak impurity physical observables behave as if the system 
is split into two semi-infinite chains. This leads to
a conductance which vanishes with a power law in $T$ at low temperatures
  \cite {KF}. A more technical discussion is presented in section 3.

Electrons are spin one-half particles and for their description it
is necessary to include the spin degree of freedom in the model. For 
a fixed quantization axis the two spin states are denoted by
$\sigma=\uparrow,\downarrow$. The fermionic creation (annihilation)
operators $c^\dagger_{n,\pm,\sigma}
(c_{n,\pm,\sigma}  )$ carry an additional spin label as well as the 
$\hat\rho_{n,\pm,\sigma}$ and the
boson operators $b_{n,\sigma}$ which in a straightforward way
generalize Eq.\ (\ref{Boson1}). The interactions $g_\nu(k)$ with $\nu=2,4$ in
Eq.\ (\ref{TL1})  become matrices $g^{\sigma \sigma '}_\nu$ in the spin labels. 
If they have the form $g^{\sigma \sigma' }_\nu(k)
=\delta_{\sigma,\sigma'}g_{\nu\Vert}(k) +
\delta_{\sigma,-\sigma'}g_{\nu\perp}(k) $ 
 it is useful 
 to switch to new boson operators $b_{n,a}$ with $a = c, s$
\begin{eqnarray}
b_{n,c} & \equiv &\frac{1}{\sqrt{2}}~ (b_{n,\uparrow} + b_{n,\downarrow})
\nonumber \\
b_{n,s} & \equiv & \frac{1}{\sqrt{2}}~ (b_{n \uparrow} - b_{n,
  \downarrow})~~,
\label{Boson2}
\end{eqnarray}
which obey $\left[ b_{a,n}, b_{a', n'} \right] = 0~$and~$ \left[
  b_{a,n}, b_{a', n'}^\dagger  \right] =  \delta_{a a'} \delta_{n n'}
  {\hat 1}$. The kinetic energy can be expressed in terms of
  ``charge'' ($c$) and ``spin'' ($s$) boson operators using
  $b^\dagger_{n,\uparrow} b_{n,\uparrow} + b^\dagger_{n \downarrow}
  b_{n \downarrow} = b_{n,c}^\dagger b_{n,c} + b_{n,s}^\dagger
  b_{n,s}$. If one defines the interaction matrix elements
  $g_{\nu,a}(q)$ via
\begin{eqnarray} 
g_{\nu,c}(q) &  \equiv g_{\nu\Vert}(q)+g_{\nu\perp}(q)\nonumber \\
g_{\nu,s}(q) &  \equiv g_{\nu\Vert}(q)-g_{\nu\perp}(q)  ~,
\label{KK}
\end{eqnarray}
and defines $\tilde {\cal N}_{\alpha,c(s)}\equiv
 (\tilde{\cal N}_{\alpha,\uparrow}
 \pm \tilde {\cal N}_{\alpha,\downarrow})/\sqrt 2
$ one can write the TL-Hamiltonian $\tilde H^{(1/2)}_{TL}$ 
for spin one-half fermions as
\begin{equation}
\tilde H^{(1/2)}_{TL} = \tilde H_{TL,c} + \tilde H_{TL,s}~,
\label{TL2}
\end{equation}
where the $\tilde H_{TL,a}$ are of the form Eq.\ (\ref{TL1})  but
 the interaction
matrix elements have the additional label $a$. The two terms on the
rhs of Eq.\ (\ref{TL2})  {\em commute}, i.e. the ``charge'' and ``spin''
excitation are completely independent. This is usually called
``spin-charge separation''. The ``diagonalization'' of the two
separate parts proceeds exactly as before and the low energy
excitations are ``massless bosons'' $\omega_{n,a} \approx v_a |k_n|$ with
the {\em charge velocity} $v_c=(v_{J_c}v_{N_c})^{1/2}$ and the {\em spin
  velocity} $v_s= (v_{J_s}v_{N_s})^{1/2}  $. The corresponding
two stiffness constants are given by  $K_c=(v_{J_c}/v_{N_c})^{1/2}$
and $K_s= (v_{J_s}/v_{N_s})^{1/2}  $.
Because of Eq.\ (\ref{TL2})
 the dependence of the 
velocities on the interaction strength (\ref{KK})
 is obtained using the results
for the spinless model following Eq.\ (\ref{Koeff}).   \\

 The low temperature
thermodynamic properties 
of the TL model
including spin, Eqs.\ (\ref{TL1})  and (\ref{TL2}),  
can be expressed in terms of the
 four velocities $v_{N_c}, v_{J_c}$, $v_{N_s}, v_{J_s} $
or the four quantities $v_c,K_c,v_s,K_s$.
Due to spin-charge separation 
the specific heat has two additive contributions of the same form as in
Eqs.\ (\ref{Debye})
and (\ref{Pauli}).
If we denote, as usual, the proportionality factor in the linear
$T$-term by $\gamma$ one obtains
\begin{equation}
\frac{\gamma}{\gamma_0}=\frac{1}{2}\left
  (\frac{v_F}{v_c}+\frac{v_F}{v_s}\right ),
\label{SW}
\end{equation}
where $\gamma_0$ is the value in the noninteracting limit.
To calculate the spin susceptibility $\chi_s$ one adds a term 
$-h\tilde{\cal N}_s$ to $\tilde H^{(1/2)}_{TL}$. 
Then by minimizing the ground state
energy with respect to $N_s$ one obtains $\langle \tilde {\cal N}_s \rangle
\sim h/v_{N_s} $, i.e. $\chi_s$ is inversely proportional to
$v_{N_s}$.
If one denotes the spin susceptibility of the noninteracting system
by   $\chi_{s,0}$, this yields for the zero temperature susceptibility
\begin{equation}
\frac{\chi_s }{\chi_{s,0} }=\frac{v_F}{v_{N_s}}=K_s\frac{v_F}{v_{s}}.
\label{Sus}
\end{equation}
For spin rotational invariant systems one has $K_s=1$ \cite{V2}.
The zero temperature compressibilty $\kappa$ is proportional to
 $(\partial^2E_0/\partial N^2)^{-1}_L $ which using Eqs.\ (\ref{TL1})
and (\ref{TL2})
leads to
 \begin{equation}
\frac{\kappa }{\kappa_0 }=\frac{v_F}{v_{N_c}}=K_c\frac{v_F}{v_c}.
\label{Kompr}
\end{equation}
 
A simple manifestation
of spin-charge separation occurs in
 the time evolution of a localized 
perturbation of e.g. the the spin-up density. The time evolution
$\alpha_{n,a}(t)= \alpha_{n,a}  e^{-i\omega_{n,a}t}$ for $a=c,s$ implies
\begin{equation}
b_{n,a}(t)=b_{n,a}\left (c_{n,a}^2e^{-i\omega_{n,a}t}-
 s_{n,a}^2e^{i\omega_{n,a}t}\right )-b^\dagger_{-n,a}c_{n,a}s_{n,a}
\left (e^{-i\omega_{n,a}t}-e^{i\omega_{n,a}t}\right )
\label{TDO}
\end{equation}
If the initial state of the system
involves a perturbation of {\it right movers}
only,
i.e. $\langle b_{n,a}\rangle =0$ for $n<0$ and the perturbation is
sufficiently smooth  $(\langle b_{n,a}\rangle \not=0$ for $0<n\ll n_c$
only) the initial perturbation is split into four parts which move
with velocities $\pm v_c$ and $\pm v_s$ without changing the initial 
{\it shape}. If only the initial expectation values of 
the $b_{n,\uparrow}$ are different from zero one obtains
for $ \delta\langle \rho_{\uparrow}(x,0)\rangle \equiv F(x)$
using Eq.\ (\ref{Boson1}) 
\begin{equation}
\delta\langle \rho_{\uparrow}(x,t)\rangle =
\sum_a \left[ \frac{1+K_a}{4} F(x-v_at)
+\frac{1-K_a}{4} F(x+v_at) \right].
\label{Frac}
\end{equation}
For the spin rotational invariant case $K_s=1$ there is no
contribution which moves to the left with the spin velocity. Already
for the pure $g_4$-model with $K_c=1$ but $v_c\not= v_s$ ``spin-charge
separation'' of the distribution occurs. For the spinless model with 
$g_2\not= 0$ the initial distribution splits into one right- and one
left-moving part, which is often called ``charge 
fractionalization'' \cite {FiGla,Pham}.
Note that the splitting described in Eq.\ (\ref{Frac}) 
 is independent of the details
of  $F(x)$ like the corresponding total charge. An additional comment
should be made: spin-charge
separation is often described as the fact that when 
an electron is injected into the system {\it its} spin and charge
move independently with different velocities. This is very misleading
as it is a {\it collective} effect of the total system which produces 
 expectation values like in Eq.\ (\ref{Frac}).

The easiest way to understand the important manifestation of
spin-charge separation
in the momentum resolved one-particle spectral functions
 \cite {MS,V} is to make use of the bosonization of the electronic
field operators discussed in the next section.

\section{ Luttinger liquids - computational tools }
In section 2 many of the important features of LL's like
the absence of a discontinuity at the Fermi surface were presented
without giving any details how these properties are actually
determined. As the most important tool the bosonization of the field 
operators is presented in detail in this section.
 This method is then used to 
calculate correlation functions like the one-particle
Green's function and the $2k_F$-density response function.
In the second part of this section the TL model 
with additional interactions and (or)
a one particle potential with a ``backscattering'' contribution 
is discussed. The model is no longer exactly solvable and 
RG arguments play an important role \cite {So,Shankar,KF}.

\subsection{Bosonization of the field operator}

In the following a selfcontained presentation of the bosonization of
 a {\em single} fermion operator 
including a straightforward construction of the particle number changing
part (``Klein factor'') is given. We present the bosonization of the
 field operator for
the right movers described by the $c_{l,+}$ and just mention the
corresponding result for the left movers.

\noindent The starting point are the  commutation relations the $c_{l,+}$
obey for $m > 0$ 
\begin{equation}
[b_m,c_{l,+}] = - \frac{1}{\sqrt{m}}c_{l+m,+}\quad , \quad
[b^\dagger _m, c_{l,+}] = - \frac{1}{\sqrt{m}}c_{l-m,+}\quad .
\label{Komm5}
\end{equation}
If (after 
taking the limit
$m_0 \to -\infty$) one
 introduces 
the $2\pi$-periodic auxiliary field operator $\tilde \psi_+(v)$,
where $v$ later will be taken as $2\pi x/L$ 
\begin{equation}
\tilde \psi_+(v)\equiv  \sum^\infty_{l = -\infty} e^{ilv}c_{l,+}~,
\label{aux}
\end{equation}
it obeys the simple commutation relations
\begin{equation}
[b_m, \tilde \psi_+(v)] = - \frac{1}{\sqrt{m}}e^{-imv}\tilde
\psi_+(v)\; ; \; [b^\dagger_m, \tilde\psi_+ (v)] = - \frac{1}{\sqrt{m}}
e^{imv} \tilde \psi_+ (v)~.
\label{Komm6}
\end{equation}
Products of exponentials of operators {\it linear } in the boson operators
\begin{equation}
A_+ \equiv \sum_{n \neq 0} \lambda_n b^\dagger_n
\quad ; \quad B_- \equiv \sum_{n \neq 0} \mu_n b_n
\label{AB}
\end{equation}
with arbitrary constants $\lambda_n$ and $\mu_n$ obey similar 
commutation relations
\begin{equation}
[b_m, e^{A_{+}} e^{B_{-}}] = \lambda_m e^{A_{+}} e^{B_{-}}
\quad ;\quad
[b_m^\dagger,e^{A_{+}} e^{B_{-}}]  =  -\mu_m e^{A_{+}} e^{B_{-}}~,
\label{Komm7}
\end{equation}
which follow from 
$[b_m, e^{\lambda b^{\dagger}_m}]=\lambda e^{\lambda b^\dagger_m}$.
We therefore make the ansatz
\begin{equation}
\tilde \psi_+ (v) = \hat O_+(v) e^{i\phi_+^\dagger (v)}
e^{i\phi_+(v)},
\label{BFO1}
\end{equation}
where the operator $i\phi_+ (v)$ is given by \cite {H}
\begin{equation}
i\phi_+(v) = \sum^\infty_{n = 1}\frac{e^{inv}}{\sqrt{n}} b_n~.
\label{BFO2}
\end{equation}
Then the operator $\hat O_+(v) $ {\it commutes} with all the $b_m$ and
 $b^\dagger_m$.
We next
construct $\hat O_+(v)$ such that both sides of
 Eq.\ (\ref{BFO1})  yield identical
matrix elements.

\noindent As $\tilde \psi_+ (v)$ reduces the number of right movers by one, the
operator $\hat O_+(v)$
also must have this property. In order to determine $\hat O_+ (v)$ we
work with the eigenstates of the noninteracting system
( the limit $m_0 \to -\infty$ is implied and $n_F$ is an arbitrary
positive integer later related to $k_F$)
\begin{equation}
|\{m_l\}_b, \tilde N_+, \tilde N_- \rangle \equiv \prod_{l}
\frac{(b^\dagger_l)^{m_l}} {\sqrt{m_l!}} \left(
 \prod^{n_F+\tilde N_-}_{n=m_0} c^\dagger _{-n,-}\right) \left(
 \prod^{n_F+\tilde N_+}_{r=m_0} c^\dagger _{r,+}\right)|{\rm Vac}
\rangle.
\label{Zustand}
\end{equation}
It is easy to see that $\hat O_+(v)|\{0\}_b,
\tilde N_+, \tilde N_-\rangle$ has no overlap to excited states
\begin{eqnarray}
\lefteqn{\langle\{m_l\}_b, \tilde N_+ -1, \tilde N_-|\hat O_+(v)|\{0\}_b,
  \tilde N_+,\tilde N_-\rangle
\;  =} \hspace{1.5cm} \nonumber \\
& & \hspace{-1.0cm}\langle \{0\}_b, \tilde N_+-1, \tilde N_-| \prod_l
\frac{(b_l)^{m_l}}{\sqrt{m_l!}}\hat O_+(v)|\{0\}_b, \tilde N_+,
 \tilde N_-\rangle~.
\label{overlap1}
\end{eqnarray}
As $\hat{O}_+(v)$ commutes with the $b_l$ the rhs of
Eq.\ (\ref{overlap1}) 
 vanishes
unless all $m_l$ are zero. This implies
\begin{equation}
\hat O_+(v)|\{0\}_b, \tilde N_+, \tilde N_-\rangle = c_+({\tilde N_+,\tilde
  N_-},v)|\{0\}_b, \tilde N_+-1, \tilde N_-\rangle~,
\label{OP1}
\end{equation}
where $c_+({\tilde N_+,\tilde
  N_-},v)$ is a c-number.
In order to determine $c_+({\tilde{N}_+,\tilde{N}_-},v)$ we
calculate 
$\langle \{0\}_b , \tilde{N}_+ -1, \tilde{N}_- |\tilde{\psi}_+(v)|
\{0\}_b, \tilde{N}_+, \tilde{N}_- \rangle$ using Eq.\ (\ref{aux}) as well as
Eq.\ (\ref{BFO1}). In the calculation of the matrix element
 with the fermionic form
Eq.\ (\ref{aux}) we use Eq.\ (\ref{Zustand})  which yields
\begin{equation}
\langle  \{0\}_b, \tilde{N}_+ -1, \tilde{N}_- | c_{l,+} | \{0\}_b,
\tilde{N}_+, \tilde{N}_- \rangle = (-1)^{\tilde{N}_-}
 \delta_{l,n_F+\tilde{N}_+}.
\label{overlap2}
\end{equation}
The factor $(-1)^{\tilde{N}_-}$ occurs because we have to
commute $c_{l,+}$ through the product of $N_-
=-m_0+1+n_F+\tilde N_-$ fermionic operators of
the left movers if we assume $-m_0+n_F$ to be odd. We note that {\em
  no} such factor occurs for the corresponding matrix element of the
left movers. The calculation of the ground state to ground state
matrix element of $\tilde \psi_+(v)$ using Eq.\ (\ref{BFO1})  is simple as both
exponentials involving the boson operators can be replaced by the unit
operator and the matrix element is just
$c_+({\tilde{N}_+,\tilde{N}_-},v)$. The comparison therefore yields
\begin{equation}
c_+({\tilde{N}_+,\tilde{N}_-},v)=(-1)^{\tilde{N}_-}e^{iv(n_F+\tilde{N}_+)}
\label{const}
\end{equation}
and $c_-({\tilde{N}_+,\tilde{N}_-},v) = e^{-i v (n_F+\tilde{N}_-)}$. Together
with Eq.\ (\ref{Zustand})  and the definition $\tilde{\cal N}_
     \alpha \equiv {{\cal N}}_\alpha - (-m_0 + 1+n_F) {\hat 1}$
this implies
\begin{equation}
\hat{O}_+(v)e^{-i (n_F+\tilde{{\cal N}}_+) v}
(-1)^{\tilde{\cal N}_-}
 |\{0\}_b, \tilde{N}_+,\tilde{N}_- \rangle = 
|\{0\}_b, \tilde{N}_+-1, \tilde{N}_- \rangle~.
\label{OP2}
\end{equation}
If we apply the operator $\hat{O}_+ (v)e^{-i
 (n_F+ \tilde{\cal N}_+)
  v}(-1)^{\tilde{\cal N}_-} $ 
to the states in Eq.\ (\ref{Zustand})  and use
  again that $\hat{O}_+(v)$ commutes with the boson operators we obtain
\begin{equation}
\hat{O}_+(v) e^{-i (n_F+\tilde{{\cal N}}_+) v} (-1)^{\tilde{{\cal N}}_-}
|\{m_l\}_b, \tilde{N}_+, \tilde{N}_- \rangle = |\{m_l\}_b, \tilde{N}_+
-1, \tilde{N}_- \rangle~.
\label{OP3}
\end{equation}
This shows that the operator $U_+ \equiv \hat{O}_+(v) e^{-i
(n_F+ \tilde{{\cal N}}_+) v} (-1)^{\tilde{{\cal N}}_-}$ is {\em independent}
      of $v$ and given by
\begin{equation}
U_+ =~\sum_{\tilde{N}_+, \tilde{N}_-}~\sum_{\{m_l\}} |\{m_l\}_b,
\tilde{N}_+ -1, \tilde{N}_- \rangle \langle \{m_l\}_b,\tilde{N}_+,
\tilde{N}_- |~.
\label{UP}
\end{equation}
It follows immediately that $U_+$ is {\em unitary}, i.e. $U_+
U_+^\dagger  = U_+^\dagger U_+ = \hat{1}$. From Eq.\ (\ref{UP})  one can infer 
that for arbitrary
functions $f$ of the number operator $\tilde{\cal N}_+$ one has
$U_+f(\tilde{\cal N}_+) = f(\tilde{\cal N}_+ + 1) U_+$.

\noindent To summarize we have shown that

\begin{equation}
\hat{O}_+ (v) = U_+e^{i(n_F+\tilde{{\cal N}}_+) v}
  (-1)^{\tilde{{\cal N}}_-}~.
\label{OP4} 
 \end{equation}
In $\hat{O}_- (u)= U_-
 e^{-i(n_F+\tilde{{\cal N}}_-) u}  $ {\it no}
 factor
$(-1)^{\tilde{{\cal N}}_+}$ appears and therefore $\hat{O}_+ (v)$ and
$\hat{O}_-(u)$ anticommute, which is necessary to enforce
anticommutation relations between $\tilde{\psi}_+(v)$ and
$\tilde{\psi}_- (u)$.
It is an easy exercise to show that e.g. the anticommutation relations
 $[\tilde{\psi}_+(v),\tilde{\psi}_+ (u)]_+=0$ are fulfilled.
 In the calculation the properties
of $ \hat{O}_+ (v) $ as well as the factor in Eq.\ (\ref{BFO1})  involving the
boson operators enter.
 If one replaces the operators
$\hat O_\alpha(v)e^{-i\alpha v(\tilde{\cal{N}}_\alpha+n_F)}$
by ``Majorana fermions'' $\eta_\alpha$ 
which commute with the boson operators and 
obey the anticommutation
relations $[\eta_\alpha,\eta_\beta]_+=2\delta_{\alpha\beta}\hat 1$, as
often
done in the literature,
this yields  $[\tilde{\psi}_\alpha(v),\tilde{\psi}_\alpha (u)]_+=
[1-\cos{(u-v)}]  e^{i\alpha (u+v)(\tilde{\cal{N}}_\alpha+n_F)}  $,
i.e. a violation of the correct anticommutation relations.
This implies that the
$U_\alpha$ have to be properly treated.
 In many publications they are written
as $ U_\alpha=e^{i\hat \theta_\alpha} $, where 
the {\it phase operators}     $\hat \theta_\alpha $
are assumed to obey the canonical commutation relations (CCR)
$[{\tilde{\cal N}}_\alpha,\hat \theta_\alpha ]=i \hat 1$
\cite{H}. We do not use this
concept here because no phase operator
can be constructed which obeys the CCR as
an operator identity \cite {S,Klein,DS,CCR,CCR2}.

In the following we will always use the ``normal ordered'' form 
(all boson annihilation operators to the right of the creation operators)
of the bosonization formula Eqs.\ (\ref{BFO1},
\ref{BFO2}). Alternatively one introduces 
a convergence factor $e^{-n\lambda/2}$, whith $\lambda \to 0$  and works
with the Hermitian Bose fields $\Phi_\alpha(v)\equiv
\phi_\alpha(v)+\phi_\alpha^\dagger(v)$ as well as the fields 
$\Phi_+\pm \Phi_-$.
 The derivatives of the latter fields are related
to the total current and the deviation of the
total charge density from its average value \cite{AAA}.
As we work with an interaction cut-off $k_c$ the introduction of
 $\lambda$
is not necessary 
and because of the space limitation this
field-theoretic formulation is not used here.

\subsection{Calculation of correlation functions for the TL model}
\noindent In order to calculate correlation
functions of the TL model with nonzero interactions it is necessary
 to express
the field operator $\tilde \psi_+(v)$ Eq.\ (\ref{BFO1})  in terms of the
$\alpha_n,\alpha^\dagger_n$ instead of the $b_n, b^\dagger_n$
 because the former have a simple time
dependence and for the temperature dependent expectation values
one has $\langle \alpha^\dagger_m \alpha_n \rangle=\delta_{mn} n_B(\omega_n)$,
where $n_B(\omega)=1/(e^{\beta \omega}-1)$ is the Bose function.
For the ground state calculation all one needs is
$\alpha_n|\Phi_0\rangle
=0$ without using the explicit form of the interacting ground state
$ |\Phi_0\rangle  $.

For periodic boundary conditions one has
 $b_m = c_m \alpha_m + s_m \alpha^\dagger_{-m}$ where the
operators $\alpha_m$ and $\alpha^\dagger_{-m}$ commute. Therefore
$e^{i\phi_+(v)}$ (and $e^{i\phi_+^\dagger(v)}$) in Eq.\ (\ref{BFO1})  can be written
as a product of two exponentials with the annihilation operators to the
right. After once using the Baker-Hausdorff formula,
$e^{A+B}=e^Ae^Be^{-\frac{1}{2}[A,B]}$ if the operators $A$
and $B$ commute with $ [A,B] $,  in order to
complete the process of normal ordering one obtains
for the physical field operator $\psi_\alpha(x)=\tilde \psi_\alpha
(2\pi x/L)/\sqrt{L}$
for a system of finite length $L$ with periodic boundary conditions
\cite {Faktor}
\begin{equation}
\psi_\alpha (x)  =
\frac{A(L)}{\sqrt{L}}\hat O_\alpha\left(\frac{2\pi x}{L}\right)
e^{i\chi^\dagger_\alpha(x)} e^{i\chi_\alpha(x)}
\label{BFO2.5}
\end{equation}
with
\begin{equation}
 i\chi_\alpha(x) = 
\sum_{m \ne 0}\frac{\theta(\alpha m)}
{\sqrt{|m|}}\left(c_m e^{ik_mx} \alpha_m - s_m
  e^{-ik_mx} \alpha_{-m}\right)~,
\label{BFO3}
\end{equation}
 $A(L) \equiv e^{-\sum^{\infty}_{n=1} s_n^2/n}$
and $\theta(x)$ is the unit step function.\\

\noindent This is a very useful formula for the 
calculation of properties of
one-dimensional interacting fermions. 
For the special choice $s_n=s(0)e^{-n/n_c}$ where $n_c=k_c L/2\pi$
is determined by the interaction cut-off, $A(L)$ can be calculated
analytically using $\sum^\infty_{n=1}z^n/n=-\log{(1-z)}$. For $n_c\gg 1$
this yields $A(L)=(4\pi/k_cL)^{s^2(0)}$ which shows
that  the prefactor in Eq.(\ref{BFO3})  has an anomalous
power law proportional to $(1/L)^{\frac{1}{2}+s^2(0)}$.
 This implies that the $c_{n,\alpha}$
scale like $(1/L)^{s^2(0)}$.
 
\noindent The time dependent operator
$\psi_+(x,t)$ follows from Eq.\ (\ref{BFO3})  by replacing $\alpha_m$ and
$\alpha_{-m}$ by $\alpha_m e^{-i\omega_mt}$ and
$\alpha_{-m}e^{-i\omega_mt}$ and $U_+$ in $\hat O_+$ by
$U_+(t)$. Various kinds of time dependent correlation functions can
quite simply be calculated using this result. Here
we begin with $iG_+^<(x,t)
\equiv \langle \psi^\dagger_+(0,0)\psi_+(x,t)\rangle$.

\noindent As $U_+$ commutes with the bosonic operator the
particle number changing operators lead to
a simple time dependent factor
\begin{equation}
U^\dagger_+ U_+(t)|\Phi_0 (\tilde N_+, \tilde N_-)\rangle
= e^{-i\left [E_0(\tilde N_+,\tilde N_-) - E_0(\tilde N_+ - 1, \tilde
  N_-)\right ]t}|\Phi_0 (\tilde N_+, \tilde N_-)\rangle.
\label{UUK}
\end{equation}
As $\psi_+(x)$ in Eq.(\ref{BFO2.5})  is normal ordered in the $\alpha$'s one has
to use the Baker-Hausdorff formula only once to normal order
$\psi^\dagger_+(0,0) \psi_+(x,t)$. This yields with $k_F=2\pi n_F/L$
\begin{eqnarray}
 ie^{i\mu t}G^<_+(x,t) 
&=&
\frac{A^2(L)}{L}e^{ik_F x} 
e^{[\chi_+(0,0),\chi_+^\dagger (x,t)]}\\ \nonumber
&=&
\frac{e^{ik_F x} }{L} e^{ \sum^\infty_{n=1}  
\frac {1}{n} \left [e^{-i(k_n x-\omega_n t)}+
2s_n^2\left (\cos{(k_n x)} e^{i\omega_n t}-1 \right) \right ]}
\label{GK1}
\end{eqnarray}
where $\mu \equiv E_0(\tilde N_+, \tilde N_-) - E_0(\tilde N_+ -1,
\tilde N_-)$ is the chemical potential.
The analytical evaluation of the sum (integral in the limit $L\to \infty$)
in the exponent in Eq.\ (\ref{GK1})  is not possible.
 An approximation which gives the correct large
$x$ and $t$ behaviour \cite {Meden} is to replace $\omega_n$ by $v_c
k_n$.
 This 
yields for $L\to \infty$  with the exponential cut-off 
for the $s_n$ used earlier \cite {LP}
\begin{equation}
ie^{i\mu t}G^<_+(x,t)=\frac{-i}{2\pi}\frac{e^{ik_F x}}{x-v_ct-i0}
\left [ \frac{r^2}{(x-v_ct-ir )(x+v_ct+ir ) }\right ]^{s^2(0)},
\label{GK2}
\end{equation}
where $r=2/k_c$. As  $\langle \psi^\dagger_+ (0,0)\psi_+ (x,0)\rangle$ for
 large $x$
decays proportional to $(1/x)^{1+2s^2(0)}$ the {\it anomalous
  dimension} for the spinless model is given by
\begin{equation}
 \alpha_L=2s^2(0)=(K-1)^2/2K .
\label{anomaleD}
\end{equation}
 Luttinger's result for 
$\langle n_{k,+} \rangle $ follows by performing the  Fourier transform 
 with respect
to $x$. The full line in Fig.\ref{Fig.Lutt} was calculated with
$s_n^2=0.3e^{-2k_n/k_c}$, while the dashed curve corresponds to 
$s_n^2=0.6(k_n/k_c)e^{-2k_n/k_c}$.
 The latter example corresponds to an interaction
with $g_2(k\to 0)\to 0$ which leads to a vanishing
anomalous dimension $\alpha_L$. In this case the occupancies $\langle n_{k,+}
\rangle $
  have a discontinuity at $k_F$ as in a Fermi liquid \cite {Diss}.
 An efficient numerical algorithm to
calculate $\langle n_{k,+} \rangle $ for arbitrary forms of $s^2_n$
is described in the appendix of reference \cite{SM}.\\

The spectral function $\rho^<(k,\omega)$ relevant for describing
angular resolved photoemission is obtained from Eq.\ (\ref{GK2}) 
by a {\it double} Fourier transform 
\begin{eqnarray}
\rho^<_+(k,\omega)
&=&
\langle
c^\dagger_{k,+}\delta[\omega+(H-E_0({\tilde N_+-1,\tilde N_-}))]c_{k,+}
\rangle \\\nonumber
&=&
\frac{1}{2\pi}\int^{\infty}_{-\infty}dt e^{i\omega t}
\int^\infty_{-\infty}dx e^{-ikx}ie^{i\mu t}G^<_+(x,t).
\label{RHOK1}
\end{eqnarray}
As Eq.\ (\ref{GK2})  is reliable in the large $x$ and $t$ limit its use in
Eq.\ (\ref{RHOK1})  correctly describes the spectral function for 
$k\approx k_F$ and $\omega\approx 0$ \cite{nonuniversal}.
 Using the variable substitutions 
$u_\mp=x\mp v_c t$ the double integral factorizes and with the additional 
approximation $i0\to ir$ 
on the rhs of Eq.\ (\ref{GK2})  one obtains \cite {Th,LP}
\begin{equation}
\rho^<_+(k_F+\tilde k,\omega) \sim \theta (-\omega-v_c|\tilde k|)
(-\omega+\tilde kv_c)^{\frac{\alpha_L}{2} -1}(-\omega-\tilde kv_c )^{
\frac{\alpha_L}{2}}
e^{r\omega/v_c}.
\label{RHOK2}
\end{equation} 
Without the additional approximation there is an additional weak dependence
on $\omega + \tilde kv_c $ \cite {MS}. The complete spectral function
$\rho_+(k,\omega)= \rho^<_+(k,\omega)+ \rho^>_+(k,\omega) $,
where $\rho^>_+(k,\omega) $ is defined via $ iG_+^>(x,t)
\equiv \langle \psi_+(x,t)\psi^\dagger_+(0,0)  \rangle$  can be
obtained using $ \rho^>_+(k_F+\tilde k,\omega)= \rho^<_+(k_F-\tilde k,-\omega)$
which follows from the particle-hole symmetry of the model.
The absence of a sharp quasi-particle peak is manifest
from $ \rho_+(k_F,\omega)\sim \alpha_L
|\omega|^{\alpha_L-1}e^{-r|\omega|/v_c}$.\\

\begin{figure} [hbt]
\begin{center}
\epsfig{file=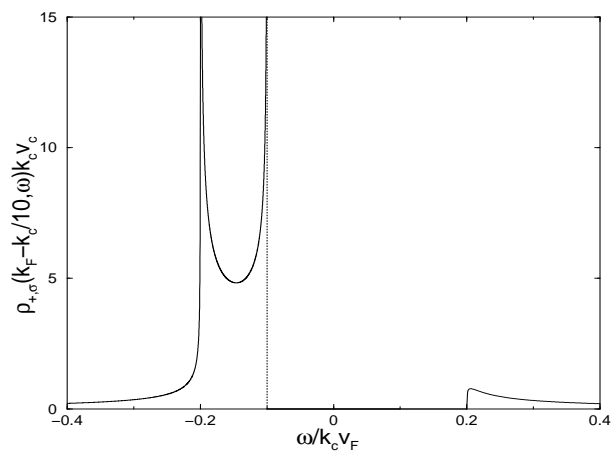,width=6cm,height=8cm,angle=-90}
\caption{Spectral function   $\rho_{+,\sigma}(k_F+\tilde k,\omega)$
 as a function
of normalized frequency for $\tilde k =-k_c/10$
for  the TL- model with  a spin independent interaction.
 The parameters are chosen such that $v_c=2v_F$ and $\alpha_L$=1/8.}
\end{center}
\label{Fig.3}
\end{figure}

 In order to calculate correlation functions of the 
spin one-half TL model the operators $b_{n,\sigma}$ which appear 
in the generalization of the 
bosonization formula Eqs.\ (\ref{BFO1})  and (\ref{BFO2})
  have to be replaced by the 
spin and charge bosons $b_{n,\sigma}=(b_{n,c}+\sigma b_{n,s})/\sqrt{2}$.
Because of the exponential occurence of the boson operators
in Eq.\ (\ref{BFO1}) 
{\it and} spin-charge separation Eq.(\ref{TL2}) the Green's function
$G^<_{+,\sigma}(x,t)$ {\it factorizes} into a spin and a charge part,
which both are of the form as the square root of the function on the 
rhs of Eq.\ (\ref{GK2}) . This square root results from the factors
$1/\sqrt{2}$ in the expression for the $b_{n,\sigma}$.
In the spin factor the charge velocity $v_c$ is replaced by
the spin velocity $v_s$. For the average occupation numbers one again
obtains Luttinger's result Eq.\ (\ref{nvk})  with $\alpha_L=s^2_c(0)+s^2_s(0)
\equiv \alpha_c+\alpha_s$. The individual contributions can be expressed  
in terms of the $K_a\equiv (v_{J,a}/v_{N,a})^{1/2}$ as $\alpha_a=
(K_a-1)^2/(4K_a)$.
As in the spinless model the fermionic (creation) annihilation
operators $c^{(\dagger)}_{n,\alpha,\sigma}$ scale like $(1/L)^{\alpha_L/2}$.
 For spin rotational invariant systems one has
$K_s=1$, 
i.e. {\it no} contribution to the anomalous dimension $\alpha_L$ from the spin 
part of the Hamiltonian \cite {V2}.
 For the momentum integrated spectral functions
one obtains 
$\rho_{\alpha,\sigma}(\omega)\sim |\omega|^{\alpha_L}$ as in the
spinless model \cite {Kommentar2}.
 The $k$-resolved spectral functions
 $\rho_{\alpha,\sigma}(k,\omega)$ on the other hand show a drastic
 difference to the model without spin. The delta peaks of the 
noninteracting model are broadened into {\it one} power law threshold
Eq.\ (\ref{GK2})  in the model without spin and {\it two}
 power law singularities (see Fig. \ref{Fig.3})
in the model including spin \cite {MS,V,Meden} (for $\alpha_L<1/2 $
in the case of a spin independent interaction). The ``peaks'' disperse
{\it linearly} with $k-k_F$.

It is also straightforward to calculate various response functions for
the TL model. We discuss the 
 density response function $R(q,z)
\equiv -\langle\langle \hat \rho_q;\hat \rho_{-q}\rangle\rangle_z/L $ of the
spinless model for $q\approx 0$
and $q\approx \pm 2k_F$, where 
\begin{equation}
\langle\langle \hat A;\hat B \rangle\rangle_z
\equiv  -\frac{i}{\hbar} \int_0^\infty \langle \left[ A(t),B\right ]\rangle
e^{izt}dt
\label{retKomm}
\end{equation}
involves the retarded commutator \cite {Z}
and $z$ is a frequency above the real axis.
 From the decomposition \cite {low}
$\psi(x)\approx \psi_+(x) +\psi_-(x)$ of the field operator $\psi(x) $
in the original Tomonaga model it is obvious that the operator
$\hat \rho (x)=\psi^\dagger(x)\psi(x)$ of the density
 (see Eq.\ (\ref{Dichte}) )
 has two very different
contributions
\begin{eqnarray}
\hat \rho(x)
&\approx&
\hat \rho_+(x) +\hat \rho_-(x)+\left (
  \psi^\dagger_+(x)\psi_-(x)+h.c.\right )\\ \nonumber
&\equiv&
\hat \rho_0(x)+\hat \rho_{2k_F}(x).
\label{RHOzerl}
\end{eqnarray}
The spatial Fourier transform of $\hat\rho_0$ is {\it linear} in the boson
operators Eq.\ (\ref{Boson1})  and the  $q\approx 0$ contribution of the density
response function $ [R(q,z)]_0 $ defined with the operators $(\hat\rho_0)_q$
 follows using the (linear) equations of motion 
for the $b_n(t)$ as
\begin{equation}
[R(q,z)]_0=\frac{1}{\pi\hbar}\frac{q^2v_J(q)}{[qv_c(q)]^2-z^2}
\label{R1}
\end{equation} 
This exact result for the     $q\approx 0$ contribution agrees with
the RPA result for the original Tomonaga model. This fact,
not mentioned in Tomonaga's paper \cite {T} 
as the RPA paper by Bohm and Pines \cite {Bohm} was not yet published, was 
``discovered'' many times in the literature.
For the spin $1/2$-model $[R(q,z)]_0$ has an additional factor
$2$ and one has to replace $v_J $ by  $v_{J_c} $.

The real part of the ($q=0$) frequency dependent conductivity
$\sigma (\omega+i0)  $
 follows from $ [R(q,\omega+i0)]_0  $ by multiplication with
$ie^2\omega/q^2$ and taking the limit $q \to 0$.
This yields for the spinless model 
\begin{equation}
(\hbar/e^2){\mbox
  Re}\sigma (\omega+i0)=v_J\delta(\omega)=K v_c\delta(\omega)
\label{Sigma}
\end{equation}
 For the Galilei invariant Tomonaga
model Eq.\ (\ref{T1})  one has $v_J=v_F$, i.e. the weight $D$ of the 
zero frequency ``Drude
peak'' is
independent of the interaction, as expected. As $D$ apart from a
constant is given by the second derivative of $E_0(\Phi)/L$ with
respect to the magnetic flux through the $1d$ ring \cite {Kohn}, $K$
(or $K_c$) can
be obtained from a {\it ground state calculation} for microscopic
lattice models  
using $K_{(c)}=(D\kappa/D_0\kappa_0)^{1/2}$, where
$\kappa$ is the compressibility discussed in Eq.\ (\ref{Kompr}).
The anomalous decay of the correlation functions for these models,
which are more difficult to calculate directly, can
then be quantitatively predicted 
if Haldane's LL concept is taken for granted. For a {\it weak}
two-body interaction the result for     $K_{(c)}-1$ {\it linear} in 
the interaction follows from first order perturbation theory
for the ground-state energy, which involves the (non-selfconsistent)
Hartree and Fock terms.
As they are independent of the magnetic flux, $D/D_0$ has {\it no} 
term linear in $\tilde v$,
 i.e. $K_c\approx (\kappa/\kappa_0)^{1/2}=(v_F/v_{N_c})^{1/2}$,
which holds exactly for Galilei invariant continuum models \cite {H2}.
Performing the second derivative of $E^{(1)}_0(N)$ with respect to 
$N$ yields \cite {local}
\begin{equation}
K_c=1-\frac{2\tilde v(0)-\tilde v(2k_F)}{2\pi \hbar v_F}
+O(\tilde v^2).
\label{KC1}
\end{equation}
In the spinless case the factor $2$ in front of $\tilde v(0) $ is
missing in the result for $K$. Instead of $D$ as the second input
besides $\kappa$ one can obtain $v_c$ directly by calculating
the lowest charge excitation energy (see section 4).
 
The easiest way to calculate the  $q\approx \pm 2k_F$ contribution
to the density response is to
use the bosonization of the field operators \cite {LP}. The first step
is to normal order $  
  \psi^\dagger_+(x)\psi_-(x) $ using Eq.\ (\ref{BFO2.5}) . 
This gives a factor $e^{[\chi_+(x),\chi^\dagger_-(x)]}$
which  using
$[\chi_+(x),\chi^\dagger_-(x)]
=-2\sum_{m>0}c_ms_m/m $     together with the factor $A^2(L)$
leads to
\begin{eqnarray}
\psi^\dagger_+(x)\psi_-(x)      =  
       \frac{a_0}{L}       \left( \frac{4\pi}{k_cL} \right)^{K-1}
\hspace{-0.3cm}
\hat O^\dagger_+(\frac{2\pi x}{L})
\hat O_-(\frac{2\pi x}{L})
e^{-i\Delta \chi^\dagger(x)} e^{-i\Delta \chi(x)} \\ \nonumber
\label{Produkt}
\end{eqnarray}
with
\begin{eqnarray*}
\Delta \chi(x) \equiv 
\chi_+(x)-\chi_-(x)=
-i\sum_{m> 0}\sqrt{ \frac{K_m}
{m}}\left( e^{ik_mx} \alpha_m -
  e^{-ik_mx} \alpha_{-m}\right).
\end{eqnarray*}
Here $a_0$ is a dimensionless constant of order unity and the exponent
$K-1$ of the second factor on the rhs follows using $2s^2_m+2c_ms_m
=K_m-1$. The importance of this factor for impurity
scattering in Luttinger liquids was first pointed out by Mattis (1974)
\cite {Mattis} and will be discussed later. 
The calculation of the two terms of the commutator
 $\langle [    \psi^\dagger_+(x,t)\psi_-(x,t), \psi^\dagger_-(0,0)\psi_+(0,0)
]\rangle$ is then straightforward and one obtains for the spectral
function of the  $q\approx \pm  2k_F$ response function 
the power law behaviour \cite {LP}
\begin{equation}
\mbox{Im} [R(\pm 2k_F+Q,\omega)]_{2k_F}\sim \mbox{sign}(\omega)
\theta(\omega^2-v^2_c Q^2)\left
 (\frac{\omega^2-v^2_c Q^2  }{v^2_ck^2_c}\right)^{K-1}
\label{R2}
\end{equation}
The {\it static} $\pm 2k_F+Q$ response diverges proportional to
$|Q|^{2(K-1)}$ which has to be contrasted with
the logarithmic singularity in the
noninteracting case.
In the model including spin the exponent $2K-2  $
is replaced by $K_c+K_s-2$.

The pair propagator $P(q,\omega)$ resulting from the response
function for $\hat A=\psi^\dagger_+(x)\psi^\dagger_-(x)$ and 
 $\hat B=\psi_-(0)\psi_+(0)$ was found by Luther and Peschel to be
the same as the $2k_F$-density response, provided the {\it sign}
of the interaction is reversed \cite {LP}. An {\it attractive} interaction 
leads to a power law divergence in $P(q=0,\omega=0)$ as the
temperature is lowered, indicative of large {\it pairing fluctuations}.\\

\subsection{The TL model with additional interactions and perturbations}
The exact solution of the TL model essentially depends on the fact
that the numbers of right and left movers are {\it conserved}. This
symmetry can be destroyed by a one-particle potential with  
$\pm 2k_F$-Fourier components or by interaction terms which
change the individual particle numbers, like
$2k_F$-``backscattering'' terms or Umklapp-terms
for a half-filled band. With such additional terms the model is
in general no longer exactly solvable. Important insights  
about the influence of such terms have come from a (perturbational)
RG analysis \cite {So,KF}.\\

\subsubsection{Impurity in a spinless TL model }
We begin with the spinless model with an additional impurity which
is described by
\begin{equation}
\hat V_I=\int \left [V_F(x)\hat \rho_0(x)+V_B(x) \hat \rho_{2k_F}(x) \right] dx
\equiv \hat V_F +\hat V_B,
\label{Potential}
\end{equation} 
where $\hat V_F$ describes the forward and $\hat V_B$ the backward scattering
due to the impurity and the two different operators for the densities
are defined in Eq.\ (\ref{RHOzerl})  .
 As the forward scattering term is {\it linear}
in the boson operators it can be treated in an exact way. The 
backscattering term has the property $[\hat V_B,{\tilde{\cal N}}_\alpha]\ne 0$ 
and the model can no longer be solved exactly (except for $K=1/2$
and a special assumption about $V_B$, as
discussed below).
For a zero range impurity it follows directly from
 Eq.\ (\ref{Produkt})  that $\hat 
V_B$ scales as $(1/L)^K$ 
while $\tilde H_{TL}$ in Eq.\ (\ref{TL1})  scales as $1/L$.
Therefore the influence of $\hat V_B$  
depends crucially on the {\it sign} of the two-body interaction \cite
{Mattis,LuPe}.
 For {\it repulsive} interactions one has $K<1$
which shows that $\hat V_B$ is a {\it relevant} perturbation. 
 For  $K>1$, i.e.  an attractive interaction, 
$\hat V_B$ is irrelevant. A detailed RG analysis of
the problem was presented in a seminal paper by Kane and Fisher \cite
{KF}. For a zero range backscattering potential
and two-body interaction they mapped the 
problem to a {\it local} bosonic sine-Gordon model \cite {KF,AAA,Kommentar3}.
The subsequent RG analysis shows that the backscattering amplitude
scales as $\Lambda^{K-1}$ when the flow parameter $\Lambda$ is sent to zero
\cite {Thierry}, as can be anticipated from Eq.\ (\ref{Produkt}) .
 This leads to the breakdown of the
 perturbational analysis in $V_B$ for repulsive interactions.
As already mentioned in section 2 this analysis was supplemented
by a RG analysis of a weak hopping between two semi-infinite chains.
The weak hopping scales to zero like $\Lambda^{\alpha_B}$
for repulsive interactions,
where $\alpha_B=K^{-1}-1$ is the {\it boundary exponent}. It 
describes e.g. the different scaling $\rho(x,\omega)\sim
|\omega|^{\alpha_B}$
of the local spectral function near  a hard wall boundary
 of a LL \cite {KF,FG,M2S4}. 
These scaling results together with the asumption mentioned
in section 2  leads to the ``split chain scenario'' in which even
for a {\it weak } impurity the observables at low energies behave
as if the system is split into two chains with fixed boundaries at
the ends.
Within the bosonic field theory this assumption was verified by
quantum Monte Carlo calculations \cite {QMC} and the 
thermodynamic Bethe ansatz \cite {Ludwig}.

 This implies e.g. for the local density of states 
 $\rho(x,\omega)\sim |\omega|^{\alpha_B}$ for small $|\omega|$ and $x$
{\it  near the impurity}
like in a LL near a hard wall. The transmission through the impurity
vanishes near $k_F$ proportional to $\sim
|k-k_F|^{2\alpha_B}$
 which leads to a conductance $G(T)$ which vanishes with temperature $T$
in power law fashion $G(T)\sim T^{2\alpha_B}$ \cite {KF}.
  
 Additional
insight comes from the analysis for the special value $K=1/2$
\cite {KF, AAA,DS} .
For $V_B(x)=V_B\delta(x)$ the expression for $\Delta \chi(0)$
 in Eq.\ (\ref{Produkt})    
can be written in terms of new boson operators $\tilde \alpha_m
\equiv (\alpha_m-\alpha_{-m})/\sqrt{2}$. If one neglects the momentum
dependence of $K_m$ in Eq.\ (\ref{Produkt})   
and puts  $K_m=1/2$ one obtains $i\Delta \chi(0)=
\sum_{m\ge 1}\tilde \alpha_m/{\sqrt{m}}  $ as in the
bosonization of a {\it single} field operator Eqs.\ (\ref{BFO1})  and
 (\ref{BFO2}) .
It is then possible to {\it refermionize} the $K=1/2$-TL model with
a zero range impurity. Even the Klein factors can properly be handled
\cite {DS}
and one obtains a model of ``shifted noninteracting Fermi oscillators''
which can be solved exactly introducing an auxiliary {\it Majorana
  fermion} \cite {AAA,DS}. Unfortunately the local densities of states
 {\it cannot} be calculated exactly because of the complicated
nonlinear relationship between the original fermion operators and the 
fermion operators which diagonalize the shifted Fermi oscillator
problem \cite {DS}. Additional results for the transport through
a spinless LL containing one impurity were obtained by mapping the
problem onto the boundary sine-Gordon model and using its
integrability \cite{Saleur}.

In order to bridge the two regimes treated by Kane and Fisher one can 
use a fermionic RG description bearing in mind that it is perturbational
in the two-body interaction \cite{Glazman,M2S2}. It shows that the {\it long
  range} oscillatory effective impurity potential is responsible
for the ``splitting'', for site impurities as well as for
hopping impurities of arbitrary strength.
 For realistic parameters 
{\it very large} systems are needed to reach the asymptotic
open chain regime \cite{M2S2}. Hence only special mesoscopic systems, such as 
very long carbon nanotubes,  are suitable for experimentally observing
the impurity induced open boundary physics.

For a discussion of the impurity problem in the TL model including
spin see also reference \cite {Furusaki}.

\subsubsection{The TL- model with additional two-body interactions}

Tomonaga was well aware of the limitations of his approach
for more generic
  two-body interactions 
(``In  the case of force of too short
range this method fails''\cite {T}). We therefore first discuss
Tomonaga's continuum model in this
short range limit $k_c\gg k_F$ 
opposite to the one considered in section 2. Then low energy
scattering processes with momentum transfer $\approx \pm 2k_F$ are 
possible and have to be included in the theoretical description of the low
energy physics.

In the ``$g$-ology'' approach
 one linearizes
the nonrelativistic dispersion around the two Fermi points and
goes over to right- and left-movers
 as in section 2. Then the ``$2k_F$''-processes
are described by the additional interaction term 
\begin{equation}
H^{(1)}_{\rm int}=\sum_{\sigma,\sigma'}
 \int
\left ( g_{1\Vert}\delta_{ \sigma, \sigma' }+ 
 g_{1\perp}\delta_{ \sigma,- \sigma' }\right )
\psi^\dagger_{+,\sigma}(x)  \psi^\dagger_{-,\sigma'}(x) 
    \psi_{+,\sigma'}(x)   \psi_{-,\sigma}(x)   dx.
\label{Hg1}
\end{equation}
For a spin-independent two particle interaction one has
$ g_{1\Vert}= g_{1\perp}=g_1 $. For the zero range interaction assumed
in Eq.\ (\ref{Hg1})  one has to introduce band cut-offs
 to regularize the interaction term. 
 The RG flow equations for the cut-off dependent 
interactions on the one-loop level are quite simple \cite{So}.
If $s$ runs from zero to infinity in the process of integrating out 
degrees of freedom one obtains for spin-independent interactions
\begin{eqnarray}
\frac{dg_1(s)}{ds}
&=&  
-\frac{1}{\pi\hbar v_F}g^2_1(s) \\ \nonumber
\frac{dg_2(s)}{ds}
&=& 
- \frac{1}{2\pi\hbar v_F}g^2_1(s)  \nonumber
\label{RG1}
\end{eqnarray}
and $g_4$ is not renormalized.
Obviously $g_1(s)$ can be obtained from the first equation only
\begin{equation}
g_1(s)=\frac{g_1}{1+s\frac{g_1 }{\pi\hbar v_F}},
\label{RG2}
\end{equation}
where $g_1$ is the starting value.
It is easy to see that
 $g_1(s)-2g_2(s)=g_1-2g_2 $ holds by subtracting
twice the second equation from the first in Eq.\ (\ref{RG1}) .
In the following we use the notation $g^*_\nu \equiv g_\nu(s\to
\infty)$.
 Now one has to distinguish two cases:

for $ g_1\ge 0$ one renormalizes to the {\it fixed line}
$g^*_1=0, \ g^*_2=g_2-g_1/2$ and the {\it fixed point Hamiltonian is a
TL model}
which shows the generic importance of the 
TL model for repulsive interactions.
 In this case the $g_1$-interaction is called
{\it marginally irrelevant}. For the nonrelativistic 
continuum model with a spin independent 
 interaction one has 
$g^*_{2c}=2\tilde v(0)-\tilde v(2k_F)$ and $g^*_{2s}=0$
and for the stiffness constant $K_c=[(2\pi v_F+g^*_{4c}-g^*_{2c} )/
 (2\pi v_F+g^*_{4c}+g^*_{2c} )]^{1/2}
\approx 1-[2\tilde v(0)-\tilde v(2k_F) ]   /(2\pi\hbar v_F)$  and
$K_s=1$. Due to the approximations made, also here only the result
for $K_c-1$
{\it linear } in $\tilde v$ is reliable. The agreement with the direct
calculation Eq.\ (\ref{KC1})  shows explicitly
 to leading order in the interaction  
that Haldane's Luttinger liquid concept is consistent.

For $g_1<0$ the solution (\ref{RG2})  shows that $g_1(s) $ {\it diverges}
at a finite value of $s$. Long before reaching this point the
perturbational analysis breaks down and all one can say is that the
flow is towards {\it strong coupling}. In this case the $g_1$-
interaction is called {\it marginally relevant}. In order to 
to obtain an understanding of the strong coupling regime it is 
useful to bosonize the additional interaction $H^{(1)}_{\rm int}$
in Eq.\ (\ref{Hg1})  \cite {LE}. The term proportional to $g_{1\Vert}$
is of the form of a $g_{2\Vert}$-interaction and therefore bilinear
in the boson operators Eq.\ (\ref{Boson2}) . For the $g_{1\perp}$-term one
uses the bosonization of the field operators Eqs.\ (\ref{BFO1}) and 
(\ref{BFO2}) with
additional spin labels. As the  $g_{1\perp}$-term
contains field operators $\psi^\dagger_{\alpha \uparrow}(x)
         \psi_{\alpha \downarrow}(x) $ of {\it opposite} spin
it only involves ``spin bosons'' Eq.\ (\ref{Boson2}) ,
 which implies ``spin-charge
separation'' also for this model \cite {Kommentar4}. The charge part
stays trivial with massless charge bosons as the elementary
interactions.
Luther and Emery showed that for a particular value of $g_{1\Vert}$
 the $g_{1\perp}$-term can be written as a product
of spinless fermion field operators
and  the exact solution for the spin part of
 the Hamiltonian is possible using {\it refermionization} \cite{LE},
 discussed earlier
in connection with the backscattering impurity.
The diagonalization of the resulting problem of noninteracting fermions 
is simple and shows that the spectrum for the spin excitations is
{\it gapped}. It is generally believed that these properties 
of {\it Luther-Emery phases} are not restricted to the
solvable parameter values.

Strong coupling phenomena which lead to deviations from LL-properties
with gapped phases are discussed in detail in section 4 for
{\it lattice models}. There in case of {\it commensurate}
filling {\it Umklapp
processes} can become important, e.g. for half filling where
 two left movers from the
vicinity of the left Fermi point are scattered into two right movers
near the right Fermi point or vice versa.
 As $G=4k_F$ is a reciprocal lattice vector
such a scattering process is a low energy process conserving quasi-momentum.
In the $g$-ology model such processes are described by an
additional term 
\begin{eqnarray}
  H^{(3)}_{\rm int}&=& \frac{1}{2}
\sum_{\sigma,\sigma '}\int g^{\sigma,\sigma '}_3(x-y)
\left [\psi^\dagger_{+,\sigma}(x)  \psi^\dagger_{+,\sigma'}(y)
    \psi_{-,\sigma'}(y)   \psi_{-,\sigma}(x)e^{2ik_F(x+y)}\right. \nonumber\\
  && \left.  \hspace{3.5cm}
+H.c.\right ]dxdy \hspace{-1.0cm}
\label{H3}
\end{eqnarray}
Umklapp processes for $ \sigma=\sigma ' $ are only possible for
nonzero interaction range.

\section{Results for integrable lattice models}

As mentioned in subsection 2.4, results for integrable models which
can be solved exactly by the Bethe ansatz played a central role 
in the emergence of the general ``Luttinger liquid'' concept \cite {H1}. 
It is therefore appropriate to shortly present results for the two 
most important lattice models of this type, the model of spinless
fermions with nearest neighbour interaction and the $1d$-Hubbard
model.
(We put $\hbar=1$ in this section.)

\subsection{Spinless fermions with nearest neighbour interaction}

The one-dimensional single band lattice model of spinless fermions
 with nearest
neighbour hopping matrix element $t(>0)$, 
and nearest neighbour interaction $U$ (often called $V$
in the literature) is given by
\begin{eqnarray}
H = -t \sum_{j} \left( c_j^{\dag} c_{j+1}^{} 
  +H.c.  \right)
  + U \sum_{j} \hat n_j \hat n_{j+1} \equiv \hat T +\hat U,
\label{spinlessferm}
\end{eqnarray}
where $j$ denotes the sites and the $\hat n_j=c^\dagger_jc_j$ are
the local occupation number operators.
 In the noninteracting limit
$U=0$ one obtains for lattice constant $a=1$
 the well known dispersion $\epsilon_k=-2t\cos k$.
For the following discussion of the interacting model ($U\ne 0$) we mainly
focus on the {\it half filled} band case $k_F=\pi/2$ with
 $v_F=2t$. In contrast to the (continuum) Tomonaga model Umklapp
terms appear when the interaction term in Eq.\ (\ref{spinlessferm})
is written in the $k$-representation. As discussed below they are
{\it irrelevant} at the noninteracting ($U=0$) fixed point \cite
{Shankar}.
 Therefore
the system is a {\it Luttinger liquid} for small enough values of $|U|$.   
The large $U$ limit of the model is easy to understand:
 For $U\gg t$ charge density
wave (CDW) order develops in which only every other site is occupied thereby
avoiding the ``Coulomb penalty''. For large but negative $U$ the
fermions want to be as close as possible and phase separation occurs.
For the quantitative analysis it is useful that the model in
 Eq.\ (\ref{spinlessferm}) can be exactly mapped to a 
$S=1/2$-Heisenberg chain
with uniaxially anisotropic nearest neighbour exchange (``$XXZ$''
model)
in a magnetic field
by use of the Jordan-Wigner transformation \cite{JW}. For $U>0$ this
model is also called the antiferromagnetic Heisenberg-Ising model.
The point $U\equiv U_c=2t$ corresponds to the {\it isotropic} Heisenberg
model. For $U>2t$ the Ising term dominates and the ground state is a
well defined doublet separated by a gap from the continuum and long
range antiferromagnetic order exists. 
For $-2t<U\leq 2t$ there is no long
range magnetic order and the spin-excitation spectrum is a gapless
continuum. The mapping to the $XXZ$-model therefore suggests that
the spinlesss fermion model Eq.\ (\ref{spinlessferm}) in the half
filled band case is a Luttinger liquid for $|U|<2t$. 

Before we present the exact results for the Luttinger liquid
parameters $K$ and $v_c$ from the Bethe ansatz solution 
\cite {H1,YY}, 
we shortly discuss the RG approach to the model. A perturbative RG
calculation around the free fermion fixed point is discussed in
detail in Shankar's review article \cite {Shankar}. The first step
is to write the four fermion matrix elements 
of the interaction $\hat U$ in  Eq.\ (\ref{spinlessferm}) in the
$k$-representation. This yields for a chain of $N$ sites with periodic 
boundary condition and values $k_j=2\pi j/N$ in the first Brillouin
zone
\begin{equation}
\langle k_1,k_2|\hat U|k_3,k_4\rangle  
= \frac{2U\cos (k_1-k_3)}{N}
\sum_{m=0,\pm 1}\delta_{k_1+k_2,k_3+k_4+2\pi m} 
\label{Umklapp}
\end{equation}
The $m=0$ term on the rhs of Eq.\ (\ref{Umklapp}) represents
the direct scattering terms and the $m=\pm 1$
terms the Umklapp processes. 
The matrix element antisymmtrized in $k_3$ and $k_4$
is proportional to $\sin{[(k_1-k_2)/2]} \sin{[(k_3-k_4)/2]}  $.
Therefore the low energy Umklapp
Hamiltonian scales like $(1/L)^3$  which shows that 
it is strongly irrelevant at the free field fixed point
\cite {Shankar}.
 This analysis confirms the Luttinger liquid
behaviour for small values of $U$, but gives no hint about the
critical value $U_c$ for the CDW transition.
With the separation $\hat U \equiv \hat U_0 +\hat U_{\rm Umklapp}$ implied
by Eq.\ (\ref{Umklapp}) one can do better by first treating $\hat T+\hat
U_0$ by bosonization and then perform the RG analysis around the 
corresponding TL fixed point to get information for which value of $U$
the Umklapp term starts to be a relevant perturbation.
For this analysis it is easier to work directly with the unsymmetrized 
matrix elements in Eq.\ (\ref{Umklapp}). As $k_1-k_3\approx \pm \pi$
for the low energy Umklapp processes this leads after extending the 
(linearized) dispersion of the right and left movers from $-\infty$
to $\infty$ to a $g_3$-interaction with a range of order $r= a$.
The scaling dimension of the corresponding $H_{\rm int}^{(3)}$ 
follows using
 bosonic normal ordering as in Eq.\ (\ref{Produkt}).
For $x-y$ of order $r$ or smaller one obtains
\begin{eqnarray}
\psi^\dagger_+(x)
\psi^\dagger_+(y)\psi_-(y)\psi_-(x)L^2
& \sim &
\left (\frac{x-y}{L}\right)^2 
\left (\frac{r}{L}\right )^{4(K-1)}\\
&&  \times (U^\dagger_+)^2U_-^2  e^{2ik_F(x+y)}
e^{i B^\dagger(x,y)}e^{iB(x,y)}~,  \nonumber
\label{Dimension}
\end{eqnarray}
where $B(x,y)=\chi_-(x)+ \chi_-(y) - \chi_+(x)-\chi_+(y) $
with $\chi_\alpha(x)$ defined in Eq.\ (\ref{BFO3}).
The first factor on the rhs is due to the Pauli principle 
and describes the same physics as the two
$\sin$-factors mentioned above for small arguments.
Therefore the second factor has to provide more than two powers
of $L$  
to make the Umklapp term a relevant perturbation, which
happens for 
$K<1/2$. As discussed below,
 the exact Bethe ansatz result for $K$
yields $U_c=2t$. If one uses the simple {\it linear} approximation
for $K-1$ in Eq.\ (\ref{KC1}) 
one obtains with Eq.\ (\ref{Umklapp})
$K^{\rm lin}=1-U/(\pi t)$ for the critical value $U^{\rm lin}_c/t=\pi /2  $,
not too bad an approximation. 

\begin{figure} [hbt]
\begin{center}
\epsfig{file=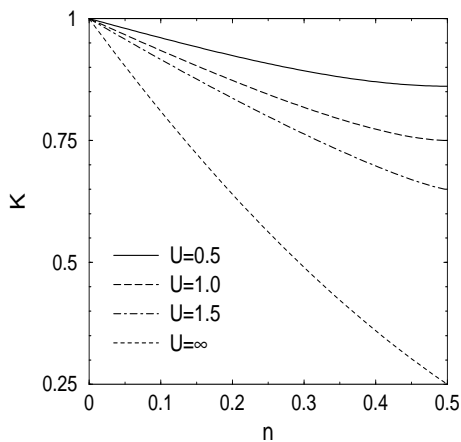,width=6cm,height=6cm,angle=-90}
\caption{Luttinger liquid parameter $K$ from the Bethe ansatz
solution as a function of the band
filling $n$ for different values of $U$  $(t=1)$.
The short dashed curve shows the infinite $U$ result $(1/2+|n-1/2|)^2$. }
\end{center}
\label{Fig.4}
\end{figure}

Exact analytical results for the Luttinger liquid parameters for the 
half filled model can be obtained from the Bethe Ansatz solution
\cite {H1,YY,Fowler}.
 It is {\it not} necessarary to address the anomalous decay 
of the correlation functions directly, but one can use a ground state
property and the lowest charge excitation to extract the parameters,
as was dicussed in connection with Eq.\ (\ref{KC1}). This yields for the
stiffness constant $  K=\pi/[2
 \arccos{ \left( - U/2t \right)}]   $ and for the
charge velocity $v_c= \pi t\sqrt{1-(U/2t)^2}
/[\pi- \arccos{ \left( - U/2t \right)}]$. 
For repulsive interactions $U>0$ the value of $K$ decreases monotonously 
from the noninteracting value
$K=1$ 
to $K=1/2$ for  $U=2t$, which corresponds to an anomalous dimension
$\alpha_L=(K+1/K)/2-1=1/4$.
 For attractive interactions $K$ diverges
when $U$ approaches $-2t$, and the charge velocity $v_c$ goes to
zero. Results for the Luttinger liquid parameter $K$ for less than
half filled bands are shown in Fig. \ref{Fig.4} \cite{Meden2}.  
The limit $a \to 0$ and $ n \to 0$ corresponds to the continuum limit. As the
interaction goes over to a contact interaction its effect vanishes
because of the Pauli principle and $K$ goes to $1$. For small enough
values of $U$
the linear approximation Eq.\ (\ref{KC1}) $K^{\rm lin}=1-U\sin{(n\pi)}/\pi t$ 
provides a good approximation for {\it all} values of $n$, in contrast to
the Hubbard model discussed below. In the infinite $U$ limit the Bethe
Ansatz equations simplify considerably and the ground-state energy as
well as low lying excited states can be calculated analytically
\cite{Fowler}.
With these results it is easy to show that $K=(1-n)^2$ holds
for $0<n<1/2$,
i.e. $K=1/4$, is the lower bound for $K$
in the LL regime of the model \cite{H1}. The
corresponding upper bound of the anomalous dimension is $\alpha_L=9/8$. 
 In order to achieve larger values of $\alpha_L$ the model in
Eq.\ (\ref{spinlessferm}) has to be generalized to include longer range
interactions \cite{CPA}.

\subsection{The Hubbard model}

As there exists an excellent review on the LL behaviour in the
$1d$-Hubbard model \cite{Schulz}, 
the following discussion will be rather short.
As the model includes spin the on-site interaction between electrons
of opposite spins is not forbidden by the Pauli principle. This is
taken as the only interaction in the model. The $1d$
Hubbard Hamiltonian
 reads
\begin{eqnarray}
H = -t \sum_{j,\sigma} \left( c_{j,\sigma}^{\dag} c_{j+1,\sigma}^{} 
  +H.c.  \right)
  + U \sum_{j} \hat n_{j,\uparrow} \hat n_{j,\downarrow} .
\label{Hubbard}
\end{eqnarray}
In the {\it extended} Hubbard model a next nearest interaction term 
$   V \sum_{j} \hat n_j \hat n_{j+1}  $ 
with $\hat n_j\equiv  \hat n_{j,\uparrow}+  \hat n_{j,\downarrow}  $
is added \cite {MilaPenc}.
 In order to show
the important difference to the spinless model Eq.\ (\ref{spinlessferm})
we again first discuss the {\it half-filled} band case, which is
metallic for $U=0$. For $U\gg t$ the ``Coulomb penalty'' is avoided
when each site is {\it singly} occpied. Then only the spin degrees
of freedom matter. In this limit the Hubbard model can be mapped
to a spin-$1/2$ Heisenberg antiferromagnet
 with an exchange coupling $J=4t^2/U$. In the charge sector there is
a large gap $\Delta_c \sim U$ while the spin excitations are gapless.
The $1d$ Hubbard model can also be solved exactly using the Bethe
ansatz \cite {LiebWu} and properties like the charge gap or
 the ground-state energy
can be obtained by solving Lieb and Wu's integral equation.
In contrast to the spinless model described in the previous subsection
the charge gap in the Hubbard model is finite for {\it all } $U>0$.
While for $U\gg t$
 it is asymptotically given by $U$  it is exponentially
small, $\Delta_c \approx (8t/\pi)\sqrt{U/t} \exp{(-2\pi t/U}$), for 
$0<U\ll t$. This shows that the Umklapp term is no longer
irrelevant at the free field fixed point. 
 The Pauli principle factor of Eq.\ (\ref{Dimension}) is 
missing as the interaction is between electrons of {\it opposite}
spin. The Umklapp term is therefore a marginal
perturbation.
The RG analysis \cite{So} shows that the Umklapp term is  marginally
{\it relevant} while the $2k_F$-backscattering (``$g_1$'') interaction
is marginally {\it irrelevant} 
for $U>0$ as discussed following Eq.\ (\ref{RG1}).

When the band is {\it not} half filled Umklapp
is not a low energy process and the Hubbard model is a Luttinger
liquid with $K_s=1$. The LL parameters $K_c$ and $v_a$ can be obtained
by (numerically) solving Lieb an Wu's integral equation \cite {Schulz2}.   
Even for $0<U\ll t$ the perturbative result Eq.\ (\ref{KC1}) works
well only for intermediate filling 
 $n\equiv N_{el}/N \approx 0.5$, where $N_{el}$ is the number
of electrons 
(half filling corresponds to $n=1$) . 
In the limit $n \to 0$ the Fermi velocity $v_F=2t\sin{(\pi n/2)}$ goes
to zero but $2\tilde v(0)-\tilde v(2k_F)=U$ stays finite and the
correction term increases with decreasing $n$ in contrast to the
spinless model. The Bethe ansatz results show that $K_c \to 1/2$ for
$n\to 0 $ as well as $n\to 1$
for {\it all} $U>0$. For $U\to \infty $ it leads to
$K_c \to 1/2$ for {\it all fillings} $n$ different from $1$.
 In this limit the velocities
are given by $v_c=2t\sin{(\pi n)}$ and $v_s=(2\pi t^2/U)[1-\sin{(2\pi
  n)}/(2\pi n)]$, i.e. the spin velocity goes to zero \cite
{Schulz,Schulz2}.
The $U=\infty$ results for $v_c$ and $K_c$ 
 can be understood {\it without} the Bethe ansatz
solution. Double occupancies of the lattice sites are forbidden and
the system behaves like a system of {\it noninteracting spinless fermions}
with $k_F$ replaced by $2k_F$ \cite {Schulz}.
 The spin degrees of freedom
play no role and any configuration of the spins gives an eigenfunction
of the same energy. 
This immediately explains the result
for $v_c$ mentioned above. For a TL model with spin 
one obtains (for fixed $N_\uparrow-N_\downarrow$)
from Eqs.\ (\ref{TL1}) and (\ref{TL2})
$L(\partial^2E_0/\partial N^2)_L=\pi v_{N_c}/2$, while the factor
$1/2$ is missing in the spinless case. The formula 
for the spinless case can be used
to calculate $ L(\partial^2E_0/\partial N^2)_L  $
for $U=\infty$ with $v_N$ replaced by $v_F(2k_F)$, using
the spinless fermion analogy. This yields $ v_{N_c}=2v_c $ i.e.
$K_c=1/2$. 

As the calculation of correlation functions not only requires excitation
energies but also many electron matrix elements which are difficult to
evaluate using the Bethe ansatz,
 various numerical methods have been used to study e.g. the
manifestation of spin-charge separation in the one-particle spectral
function \cite {Penc,Hanke}. The Bethe ansatz approach simplifies 
in the infinite $U$ limit \cite {Shiba}.
After earlier work \cite{Flo,Sumit}
the frequency dependent optical conductivity of the $1d$ Hubbard model
was also studied using Bethe ansatz methods \cite { Nunu,Fabian},
as well as the dynamical density-matrix renormalization group \cite {Fabian}.

\section{Weakly coupled chains: the Luttinger to Fermi liquid transition}

Strictly one-dimensional systems are a theoretical idealization.
Apart from this even the coupling to an experimental probe presents
a nontrivial disturbance of a Luttinger liquid. Unfortunately the weak
coupling of a $1d$ system to such a probe as well as the coupling
between several LLs is theoretically not completely
understood \cite {V2}.
The coupling between the chains in a very anisotropic $3d$ compound 
generally, at low enough temperatures, leads to true {\it long-range
  order}.
The order develops in the phase for which the algebraic
decay of the correponding correlation function
of the single chain LL is most slowly \cite {Schulz}.
This can lead e.g. to charge-density wave (CDW), spin-density wave (SDW)
order or superconductivity.
 
 In the following we shortly address some important
issues of the coupled chain problem, which are a
prerequisite for the theoretical
descriptions of the attempts to experimentally verify LL behaviour.
In the first part of this section theoretical aspects of the problem
of an infinite number of coupled chains are
addressed. This is followed by a short discussion of the 
(approximate) experimental realizations of LLs. As there are other
chapters in this book addressing this question the discussion will
be rather short.

\subsection {Theoretical models}

We consider a systems of $N_\perp$ coupled chains described
by the Hamiltonian
\begin{equation}
H=\sum_{i=1}^{N_\perp}H_i +\sum_{i\ne j}H_{ij}^{(ee)}
 +\sum_{n, (\sigma)} \sum_{i,j}t_{\perp,ij}
c_{n,
(\sigma),i}^\dagger,
 c_{n,(\sigma),j}
\label{Kettenham}
\end{equation}
where the $H_i$ are the Hamiltonians of the individual chains,
the $H_{ij}^{(ee)}$ represent the two-body (Coulomb)
 interaction of electrons on
different chains and the last term $H^{(t_\perp)}$
describes the hopping between the chains with $t_{\perp,ij}$ the
transverse hopping matrix elements and the 
$c_{n,(\sigma),i}^{(\dagger)}$ the (creation) annihilation
operators of one-particle states with quasi-momentum $k_n $
 along the chain $i$
and spin $\sigma$ (if spin is included in the model).
The individual $H_i$ can be TL-Hamiltonians Eq.\ (\ref{TL1}) or
lattice Hamiltonians like in Eqs.\ (\ref{spinlessferm}) or
(\ref {Hubbard}).   

We address the question if LL physics survives in such a model.
The second and the third term on the rhs of Eq.\ (\ref{Kettenham})
describe different types of couplings between the chains. 
If the transverse hopping is neglected ($t_\perp\equiv 0$) 
the model can be solved
exactly for special assumptions about the two-body
interaction and the $H_i$. If the individual chains are described
by TL-Hamiltonians Eq.\ (\ref{TL1}) and the interaction $H_{ij}^{(ee)}$
can be expressed in terms of the densities $\hat \rho_{n,(a),\alpha,i}$ 
the exact solution is possible by bosonization \cite{Schulz3,KMS}.  
This is important when the long range Coulomb interaction is taken 
into account. For a {\it single} chain the corresponding
one-dimensional Fourier transform $\tilde v(q)$ (which has to be
regularized at short distances) has a logarithmic singularity for
$q \to 0$. This leads to $K_{(c)}=0$ and the divergence of the anomalous
dimension, i.e. the system is {\it not } a LL. The $4k_F$ harmonic
of the density-density correlation function shows a very slow decay
almost like in a Wigner crystal \cite {SchulzWigner}.
 The Coulomb coupling
{\it between } the chains removes this singularity and a
three-dimensional extended system of coupled chains {\it is} a LL
\cite {Schulz3}. The corresponding anomalous dimension can be
calculated and leads to values of order unity for realistic 
values of 
the coupling constant $e^2/(\pi \hbar v_F)$ \cite {KMS}. If $2k_F$-scattering
terms of the interaction are kept the model can no longer be solved
exactly and a more complicated scenario emerges 
in the parquet approximation \cite{GD}.

The inclusion of the transverse hopping presents a difficult problem
even if the inter-chain two-body interactions are {\it neglected}. 
This is related to the fact that the transverse hopping is a {\it
relevant} perturbation for $\alpha_L<1$\cite {Bourbo,Wen,Bourbo1}. 
This can easily be seen if the individual chains are described
by TL-Hamiltonions Eq.\ (\ref{TL1}), scaling like $1/L$.
As discussed in section 3 the $c_{n,(\sigma),i}^{(\dagger)}$
scale like $(1/L)^{\alpha_L/2}$. As $H^{(t_\perp)}$ involves products
of creation and  annihilation operators on {\it different} chains
no further boson normal ordering is necessary and  $H^{(t_\perp)}$
scales as $(1/L)^{\alpha_L}$.
This suggests ``confinement'' for $\alpha_L>1$: if an extra electron is
put on the $j$-th chain it stays there with probability close to $1$
even in the long time limit.
This conclusion can be questioned as
RG calculations perturbative in $t_\perp$
 demonstrate that the
hopping term generates new and relevant interchain two-particle
hoppings.
These calculations show that the system flows to a strong-coupling fixed
point which cannot be determined within the approach
\cite{Bourbo1,Russen}.


If inter-chain two-body interactions are {\it included} the relevance
of hopping terms can be different. When only
 density-density and current-current interactions between the wires
are included, as discussed above \cite {Schulz3,KMS}, the possible
relevance around this Gaussian model, recently called {\it sliding} LL
   \cite{Lub,Carpent,Lub2}, can be different. 
If the single chains are in the 
spin-gapped Luther-Emery regime \cite {LE}
 single-particle hopping between the chains is
irrelevant and the coupled system can show power-law correlations
characteristic of a $1d$-LL \cite {Lub,Lub2}. For the spinless model 
single particle and pair hoppings can be irrelevant for strong enough
forward interactions \cite {Carpent}.

\begin{figure} [hbt]
\begin{center}
\epsfig{file=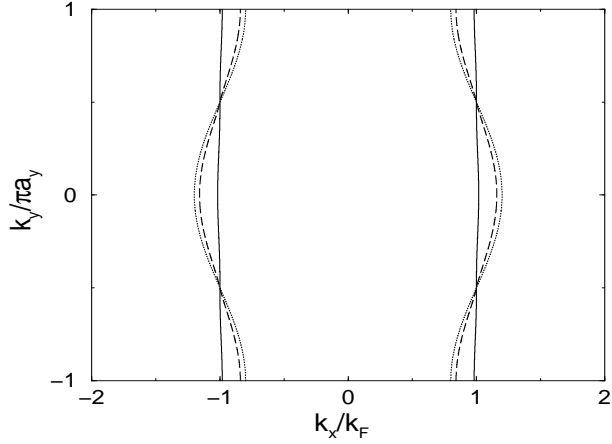,width=6cm,height=8cm,angle=-90}
\caption{Fermi surface ``flattening'' in Wen's approximation 
for coupled chains
for different values of the the anomalous dimension $\alpha_L$
for a single chain. The dotted lines show the noninteracting FS, the
long dashed curves correspond to $\alpha_L=0.125$ and the full ones to
$\alpha_L=0.6$. At  $\alpha_L=1$ the FS degenerates to two parallel
lines as without interchain coupling, called the
``confinement transition''.}
\end{center}
\label{Fig.5}
\end{figure}

In the following we concentrate on the Luttinger to
Fermi liquid crossover.
In order to get a {\it quantitative} picture
 it is desirable to study models which allow 
{\it controlled approximations}. The simple perturbative calculation
in $t_\perp$ for the calculation of the one-particle Green's function 
by Wen \cite {Wen} discussed below is unfortunately  only controlled 
in the rather unphysical limit when the transverse hopping is
independent of the distances of the chains ($t_{\perp,ij}\equiv
t_\perp$)\cite {Bourbo2}.
The (retarded) one-particle Green's function $G$ 
 is expressed in terms
of the selfenergy $\Sigma$
\begin{equation}
G(k_\Vert,\vec k_\perp,z;t_\perp)
=\frac{1}{z-\epsilon_{k_\Vert, \vec k_\perp}
-\Sigma(k_\Vert, \vec k_\perp,z;t_\perp)},  
\label{selfenergy}
\end{equation}
where $\epsilon_{k_\Vert, \vec k_\perp }$ denotes the energy dispersion for the
noninteracting model and $z=\omega +i0$ is the frequency above the
real axis. For small $t_\perp$ 
the dispersion can be linearized around
$k_\Vert=\pm k_F$ near the {\it open} noninteracting Fermi surface.
This yields
 $ \epsilon_{k_\Vert, \vec k_\perp} \approx
\pm v_F(k_\Vert \mp k_F)+t_\perp (\vec k_\perp)$.
In the context of Fermi liquid theory the selfenergy is studied
in (all orders) perturbation theory in the two-body interaction
$v$ around the noninteracting limit. This can be done using standard
Feynman diagrams. In the present context one wants to study how
the LL behaviour for {\it finite} two body interaction and finite
anomalous dimension is modified by the transverse hopping. Similar to
perturbation theory for the Hubbard model around the atomic limit
nonstandard techniques have to be used \cite {Metzner}. 
The simplest approximation, which corresponds to the ``Hubbard I''
 approximation for the Hubbard model, is to replace $\Sigma$ in
Eq.\ (\ref{selfenergy}) in zeroth order in $t_\perp$ by the selfenergy     
$\Sigma^{(\rm chain)}(k_\Vert,z)$ of a single chain \cite {Wen}. This
approximation first used by Wen reads for $k_\Vert \approx k_F$
\begin{equation}
G(k_\Vert,\vec k_\perp,z;t_\perp)_{\rm Wen}
=\frac{1}{\left [G_+(k_\Vert,z)\right ]^{-1}-t_\perp  (\vec k_\perp) },
\label{Wengl}
\end{equation}
where $G_+$ is determined by the spectral function $\rho_+$ discussed 
following Eq.\ (\ref{RHOK2}) via a Hilbert transform. In the asymptotic
low-energy regime 
this yields 
$G_+(k_F+\tilde k_\Vert,z)=A_0[(\tilde 
k_\Vert/k_c)^2-(z/\omega_c)^2]^{\alpha_L/2}
/(z-v_c\tilde k_\Vert)$ for spinless fermions, 
with $\omega_c\equiv k_cv_c$
and $A_0=\pi\alpha_L/[2\sin{(\pi\alpha_L/2)}]$. Wen's approximate
Green's function leads to a spectral function with the {\it
  same} range of continua as $ \rho_+(k_\Vert,\omega)$. In {\it addition}
there can be {\it poles} at $\omega_{ k_\Vert, \vec k_\perp }$,
determined by setting the denominator in Eq.\ (\ref{Wengl}) equal
to zero. The poles located at $\omega_{ k_\Vert, \vec k_\perp }=0$
determine the Fermi surface $\tilde k_\Vert(\vec k_\perp)$ 
of the {\it interacting} coupled system. 
From Eq.\ (\ref{Wengl}) and the simple form of $G_+$ one obtains
 $A_0(\tilde k_\Vert/k_c)^{(1-\alpha_L)}= t_\perp (\vec k_\perp)$,
which shows that the reduction of warping of the Fermi surface (FS) 
  by the interaction is proportional to
 $[t_\perp(\vec k_\perp)/\omega_c]^{\alpha_L/(1-\alpha_L)}$.
This is shown in Fig. \ref{Fig.5} for a two dimensional system of
coupled chains.
If one writes $t_\perp(\vec k_\perp)\equiv t_\perp c(\vec k_\perp )$,
with $c(\vec k_\perp) $ a dimensionless function, 
the new effective low energy scale is given by
$t_{\rm eff}=\omega_c (t_\perp/\omega_c)^{1/(1-\alpha_L)}$.
The weights $Z_{\vec k_\perp}$ of the poles for $\vec
k$ values on the Fermi surface are also  proportinal to 
 $[t_\perp(\vec k_\perp)/\omega_c]^{\alpha_L/(1-\alpha_L)}$.  
Wen's approximate solution has the Fermi liquid type property of
quasi-particle poles with nonzero weight on the Fermi surface, except
at the special points where $t_\perp(\vec k_\perp )$ vanishes.
The improved treatment by Arrigoni \cite {Arrigoni} shows that this
peculiar vanishing of the quasi-particle weigths is an artefact
of Wen's approximation. The new idea involved is to let the number
of ``perpendicular'' dimensions ``$d-1$'' go to infinity. 
This extends the original idea of the ``dynamical mean field theory''
(DMFT) \cite {MeVo}, where one treats
the Hubbard model in infinite dimensions as
 an effective {\it impurity} problem
to the case of a {\it chain} embedded in an effective medium.  
Results are obtained by carrying out a resummation of {\it all} diagrams
in the $t_\perp$-expansion which contribute in this large dimension
limit \cite {Arrigoni}.
 This approach shows explicitly how the leading order Wen
approximation is uncontrolled at low energies. For the case of
weakly coupled one-dimensional Mott insulators one expects the 
approximation to be better controlled \cite{Ess}.

 Despite the
Fermi liquid like properties at energy scales much smaller
than  $t_{\rm eff}$ the coupled chain system can nevertheless
show LL like properties for energy scales larger than $t_{\rm eff}$
if there is a large enough energy window to the high energy
cutoff $\tilde \omega_c$ which describes the regime where the asymptotic
LL power laws hold for a single chain. Then for temperatures lower than      
  $\tilde \omega_c$ but higher than  $t_{\rm eff}$ the system behaves
like a LL. The integrated spectral functions $\rho^<_{\alpha, 
(\sigma)}(\omega)$
probed by angular integrated photoemission, for example, show
approximate power law behaviour $\sim (-\omega)^{\alpha_L}$
for temperatures larger than  $t_{\rm eff}$
in the energy window $k_BT <-\omega <\tilde \omega_c$.   
Unfortunately little is known about the value of $\tilde \omega_c$
for microscopic models. An exception is the Tomonaga model Eq.\ (\ref{T1}) with
a constant $\tilde v(k)$ up to the cutoff $k_c$, where the high energy
cutoff  $\tilde \omega_c$ equals 
 $ \omega_c={\rm min}(v_c,v_s)k_c$ \cite {S}.
This implies for the integrated spectral function 
for the very large $U$ Hubbard model with periodic
boundary conditions that
the power law $|\omega|^{\alpha_L}$ only holds in a narrow energy 
window $\sim v_s$, which vanishes proportional to $1/U$ 
in the $U\to \infty$ limit \cite{MilaPenc}.
Another example is the Hubbard model at
 boundaries
 where  $\tilde
\omega_c$ can be very small for small $U$ \cite{M2S4}.

As an alternative way to treat the ``anisotropic large dimension model''
\cite {Arrigoni} one can try to solve the resulting
chain-DMFT equations numerically, using e.g. a quantum Monte Carlo
algorithm \cite {BGLG}. In this reference
the $H_i$ were chosen as Hubbard Hamiltonians
 (\ref{Hubbard}) with chain lengths 
of 16 and 32 sites. The results for a partly filled band   
as a function of temperature indicate in fact a crossover
from a LL to a FL at the estimated crossover scale 
as the temperature is lowered. In agreement with
Arrigoni \cite {Arrigoni} the authors find that the quasi-particle
weight is more uniform along the Fermi surface than suggested
by Wen's approximation Eq.\ (\ref{Wengl}). At half filling
and low but finite temperatures the crossover from  
the Mott insulator to FL crossover was examined (the intermediate
LL regime was too narrow to be visible). In the future it is to be
expected that this method applied to longer chains and additional
nearest neighbour interaction will provide important results 
which allow a more realistic comparison with experimental work.

Because of space limitations the interesting field of a {\it finite} number
of coupled chains cannot be discussed here \cite{Ketten}. 

\subsection {On the experimental verification of LL behaviour}

There exist several types of experimental systems were a predominantly
$1d$ character can be hoped to lead to an (approximate) verification
of the physics of Luttinger liquids. In the following we present a short   
list of the most promising systems and discuss some of the experimental
techniques which have been used. As these topics are also discussed in
other chapters of this book we do not attempt a complete list
of references but only refer to most recent papers or to review articles on
the subject.

\noindent The following systems look promising:
\begin{itemize}

\item Highly anisotropic ``quasi-one-dimensional'' conductors

There has been extensive work on organic conductors, like the Bechgaard
salts \cite {Jerome, Grioni}, as well as inorganic materials
\cite {Gweon,Allen}.

\item Artificial quantum wires

Two important types of realizations are quantum wires in 
{\it semiconductor heterostructures} \cite {Stormer,West}  or quantum
wires on {\it surface substrates} \cite {Yves,Himpsel}.

\item Carbon nanotubes

The long cylindrical fullerenes called quantum nanotubes
are also quantum wires  but have been
listed separately because of their special importance in future
applications like ``molecular electronics'' \cite{Dresselhaus, Dekker}.
Using the peculiar band structure of the
$\pi$-electrons of a single graphite plane
it was shown that single wall
 ``armchair'' nanotubes should show LL behaviour
with $K_c \sim 0.2 -0.3 $ down to very
low temperatures \cite {GE,KBF}, despite the fact that {\it two} 
low energy channels are present.

\item Fractional quantum Hall fluids

Electrons at the edges of a two-dimensional fractional quantum Hall
system can be described as a {\it chiral Luttinger liquid} \cite
{Wenchiral}. The power law tunneling density of states observable in
the tunneling current-voltage characteristics shows power
laws of extraordinary quality \cite {West2}.
The theoretical predictions for general filling factors
between the Laughlin states $\nu=1$ and $\nu=1/3$ \cite{FH1,FH2}
are not bourne out by experiment \cite{Mat}. 
As in these chiral LLs the right- and left-movers are {\it spatially
separated} the edge state transport is quite
different from the case of quantum wires and FQH fluids 
 are not further discussed in the following.

\end{itemize}

\noindent Promising experimental techniques to verify LL behaviour are:

\begin{itemize}

\item High resolution photoemission

One of the earliest claims of possible verification of Luttinger
liquid behaviour was from angular integrated photoemission of
the Bechgaard salt $({\rm TMTSF})_2{\rm PF}_6$, which showed a power law
supression at the chemical potential with an exponent of order $1$
over an energy range of almost one eV \cite {Yves2}. There are
serious doubts that this suppression can be simply explained by the
LL power law behaviour \cite {Grioni}. Therefore a large number of
other quasi-one-dimensional conductors were examined \cite {Grioni,
Gweon,Allen,Claessen}.
In addition periodic arrays of quantum
wires on surface substrates were studied by angular resolved
photoemisssion (ARPES), but the interpretation of a two peak structure
as spin-charge separation \cite{Yves} was questioned \cite {Himpsel}. 
Spin-charge separation was shown to occur in the $1d$ Hubbard
model also at higher energies on the scale of the conduction
band width \cite{Hanke,Shiba,Nunu}. Recent ARPES spectra of TTF-TCNQ
were interpreted with the $1d$ Hubbard model at finite
doping to show signatures of spin-charge separation over an energy
scale of the conduction band width. As for the Hubbard model 
$K_c>1/2$ for $n \ne 1$ which implies $\alpha_L <1/8$
for the anomalous dimension the experimentally found nearly linear 
spectral onset at low energies cannot be explained within the same
model. 
 ARPES data for the ``Li purple bronze'' seem
to compare favorably to the LL lineshape \cite {Allen}.
For the quasi-one-dimensional antiferromagnetic insulators ${\rm SrCuO}_2$
and ${\rm Sr}_2{\rm CuO}_3$  ARPES spectra have been interpreted
to show evidence of spin-charge separation \cite {Kim}.
For a more in depth discussion see the chapter by Grioni
in this book.
      
\item   Transport 

As discussed in section 3 even a single impurity has a drastic effect
on the conductance of a LL, which vanishes as a power law with
temperature. Another issue is the 
 ``conductance puzzle''
of a clean LL.
 There has been an extended discussion whether the 
 quantized value $e^2/h$ for noninteracting electrons
in a single channel is modified by the interaction to $K_c e^2/h$
 \cite{Maslov,Safi}. Apparently the answer depends sensitively
on the assumptions made about the contacts, a very delicate
theoretical as well as experimental problem \cite {Grabert}.
Experimental results are available for 
cleaved edge overgrowth quantum wires \cite {Stormer}
as well as carbon nanotubes \cite {Bockrath,Yao,Postma}.
In the nanotubes the authors observe approximate power laws
of the conductance which seem to be consistent with
LL behaviour.  
A detailed dicussion of transport through quantum wires
is presented in the chapter by Yacoby.  
For a recent theoretical discussion of 
experimental results on the interchain transport in the
Bechgaard salts see references \cite {Ant1,Ant2}. There
the question of
energy scales and the importance of the proximity of the incipient
Mott insulator are addressed.

\item Optical properties

Optical properties have long been used to investigate electronic 
properties of quasi-one-dimensional systems \cite {Gruener}.
The optical behaviour of different Bechgaard salts was analyzed
recently using LL concepts \cite {Degiorgi}.
At low energies, smaller than about
ten times the Mott gap, the importance of dimerization and interchain
hopping
was pointed out \cite {Fabian2}. 
 As there is a separate chapter about the optical response
in chains and ladders
 it will not be discussed further here.

\end{itemize}

\noindent Obviously neither the list of systems nor that of methods
is coming close to being complete. They were presented to show that there
are intensive experimental activities in the attempt
 to verify the elegant LL concept
put forward by theoreticians. Further work on both sides is necessary
to come to unambiguous conclusions.\\

\noindent {\bf Acknowledgements}:
For useful comments on the manuscript the author would like to thank
J. Allen, E. Arrigoni, D. Baeriswyl, L. Bartosch, J. von Delft,
R. Egger, F. Essler, F. Gebhard,
 A. Georges, T. Giamarchi,
M. Grayson, P. Kopietz, V. Meden, W. Metzner, and J. S{\'o}lyom.


\begin{thebibliography}{999}
\bibitem{Fermisee} Obviously there is the restriction $\sum_{j>0} n_j\le
  n_F$ which shows in agreement with Eq.\ (\ref{Komm2})  that the mapping is 
straightforward only for the low energy excitations which leave all
levels deep
in the Fermi sea occupied.
For noninteracting fermions the canonical partition function can also
be calaculated analytically when the energy levels have a lower bound (see
K. Sch\"onhammer, Am. J. Phys. {\bf 68}, 1033 (2000)).
\bibitem{Index} As the greek letter $\alpha$ is used as the index
to distinguish the right and left movers we write $\alpha_L$
(``Luttinger'')
for the anomalous dimension.
\bibitem{Faktor} Some authors factor out $e^{i\alpha k_F x}$
in the 
definition of the $\psi_\alpha(x)$,
which is contained in our $\hat O_\alpha(2\pi x/L)$.
\bibitem{local} It is assumed that the (quasi-) momentum matrix
elements $v_{k_1,k_2;k_3,k_4}$ of the two-body interaction
have the property   $v_{k_1,k_2;k_1,k_2}=\tilde v(0)$  and
 $v_{k_1,k_2;k_2,k_1}=\tilde v(|k_1-k_2|)$. Then it is also easy
to show directly that  $v_J=v_F+O(\tilde v^2)$
 for non 
Galilei invariant models.
\bibitem{Kommentar4} Here again the Klein factors are treated in 
a sloppy way. Strictly speaking even the fact that the Hamiltonian
is a sum of commuting boson and 
particle number parts is lost.
\end{thebibliography}
\end{document}